\documentclass[preprint,12pt]{elsarticle}

\usepackage[T1]{fontenc}
\usepackage{hyperref}
\usepackage{amsmath}
\usepackage{lineno}
\usepackage{xcolor}

\usepackage{amssymb}

\journal{NIM A}

\begin{document}

\begin{frontmatter}

\title{Precise synchronization of a free-running Rubidium atomic clock with GPS Time for applications in experimental particle physics}

\author[a]{Claire Dalmazzone\footnote{Corresponding author: claire.dalmazzone@lpnhe.in2p3.fr, +33617925827}}
\author[a]{Mathieu Guigue}
\author[a]{Lucile Mellet\footnote{Now at Michigan State University, Department of Physics and Astronomy, East Lansing, Michigan, USA}}
\author[a]{Boris Popov}
\author[a]{Stefano Russo}
\author[a]{Vincent Voisin}

\affiliation[a]{organization={Laboratoire de Physique Nucl\'eaire et des Hautes Energies (LPNHE), Sorbonne Universit\'e, CNRS/IN2P3}, addressline={4 place Jussieu}, city={Paris}, postcode={75005}, country={France}}

\author[b]{\\Michel Abgrall}
\author[b]{Baptiste Chupin}
\author[b]{Caroline B. Lim}
\author[b]{Paul-Éric Pottie}
\author[b]{Pierre Ulrich}

\affiliation[b]{organization={LNE-SYRTE, Observatoire de Paris, Universit\'e PSL, CNRS, Sorbonne Universit\'e}, addressline={61 avenue de l'Observatoire}, city={Paris}, postcode={75014}, country={France}}

\begin{abstract}

We present results of our study devoted to the development of 
a time correction algorithm needed to precisely synchronize a free-running Rubidium atomic clock with the Coordinated Universal Time (UTC). This R\&D is performed in view of the Hyper-Kamiokande (HK) experiment currently under construction in Japan, which requires a synchronization with UTC and between its different experimental sites with a precision better than $100$~ns. We use a Global Navigation Satellite System (GNSS) receiver to compare a PPS and a $10$~MHz signal, generated by a free-running Rubidium clock, to the Global Positioning System (GPS) Time signal.
We use these comparisons to correct the time series (time stamps) provided by the Rubidium clock signal. We fit the difference between Rubidium and GPS Time with polynomial functions of time over a certain integration time window to extract a correction of the Rubidium time stamps in offline or online mode. In online mode, the latest fit results are used for the correction until a new comparison to GPS Time becomes available. We show that with an integration time window of around $10^4$ seconds, we can correct the time stamps drift, caused by the frequency random walk noise and the deterministic frequency drift of the free running Rubidium clock, so that the time difference with respect to GPS Time stays within a $\pm5$~ns range in both offline or online correction mode. Presented results could be of interest for other experiments in the field of neutrino physics and multi-messenger astrophysics.\\
\end{abstract}


\end{frontmatter}

\flushbottom

%%%%%%%%%%%%%%%%%%%%%%%%%%%%%%%%%%%%%%%%%%

\section{Introduction}

A precise synchronization with the Coordinated Universal Time (UTC) or with another signal is a necessity in many applications, particularly in long-baseline physics experiments spread over several experimental sites. A good example is long-baseline neutrino oscillation experiments, like OPERA~\cite{OPERA} (2006-2012), T2K~\cite{T2K} (from 2010) and NOvA~\cite{NOvA} (from 2014), where a beam of neutrinos is produced and characterized in a first experimental site and detected, after several hundreds of kilometers of propagation, at another site to measure a change of the beam properties. Two next generation long-baseline neutrino experiments are being built at the moment: Hyper-Kamiokande (HK)~\cite{HKDesignReport} that plans to start taking data in 2027 and DUNE~\cite{DUNE,DUNE_TIMING} that should begin sometime after 2029. These experiments require a synchronization of $100$~ns or better between the different experimental sites. Moreover, multi-messenger programs that plan to compare different components of astrophysical events~\cite{Multi-messenger} (e.g.: gamma-ray bursts, gravitational waves, neutrino emissions of supernovae, etc.) require a synchronization with UTC of different experiments located all over the world. For instance, to enter the SuperNova Early Warning System (SNEWS) network~\cite{SNEWS}, a synchronization to UTC better than $100$~ns is required.

Many long-baseline physics experiments use atomic oscillators as frequency references because of their good short term stability. Among the reference oscillators available on the market, Rubidium atomic clocks are generally chosen for their affordability as it was the case for the T2K~\cite{T2KTOF} and Super-Kamiokande~\cite{SKDetector} timing systems.
 However, Rubidium clocks usually drift away from a stable reference because of frequency drift and random walk. For synchronization to UTC, this drift usually needs to be prevented or corrected. 
 
 A common solution is to discipline the average frequency of the clock to the signals of an external Global Navigation Satellite System (GNSS) receiver, with an integration time window chosen so that it does not deteriorate the short term stability of the clock. 
However, it presents some drawbacks like the fact that the user has little control on the setup. In case of problems (like jumps in the time comparison), it is difficult to understand where they come from (GPS Time, receiver, the master clock, etc.) and to assess the uncertainty on the synchronization to UTC. The R\&D work presented in this paper and introduced in~\cite{ProceedingsHK} is focused on designing and characterizing an alternative method that allows more freedom to the user and a better understanding of the process. It is based on known metrology techniques~\cite{FundaMetro,OperaTimeSyst}. The proposed method uses a free-running atomic clock to derive a time signal for time-tagging the events detected by a neutrino detector for instance. The time stamps are corrected in post-processing using comparisons of the Rubidium clock signal to GNSS Time. Indeed, the GNSS was proven to be an effective tool for clock comparison \cite{Riedel2020} and stability transfer of remote frequency standards \cite{Margolis2024}, especially when no direct fiber link is available, like for intercontinental comparison \cite{Morzynski2024} or for long baseline neutrino experiments. 

With this method, one can store all the information (the raw signal, the comparisons to GNSS Time, the derived correction etc.) and apply the correction in either online (during the data-acquisition) or offline modes. Let us note that the GNSS time is a good approximation of UTC, within a few nanoseconds, and it allows synchronization to UTC via a common-view technique~\cite{CommonView}. The common-view would be performed with a national metrology laboratory providing a local realization of UTC called UTC(k), like e.g. the NICT laboratory in Japan~\cite{NICT}.Then the conversion to UTC can be performed with the help of the Circular T of the BIPM (Bureau International des Poids et Mesures)~\cite{CircularT} at the end of each month.  

\section{Materials and Methods}

\subsection{Experimental setup}

\begin{figure}[tbp]
\centering
\includegraphics[width=\textwidth]{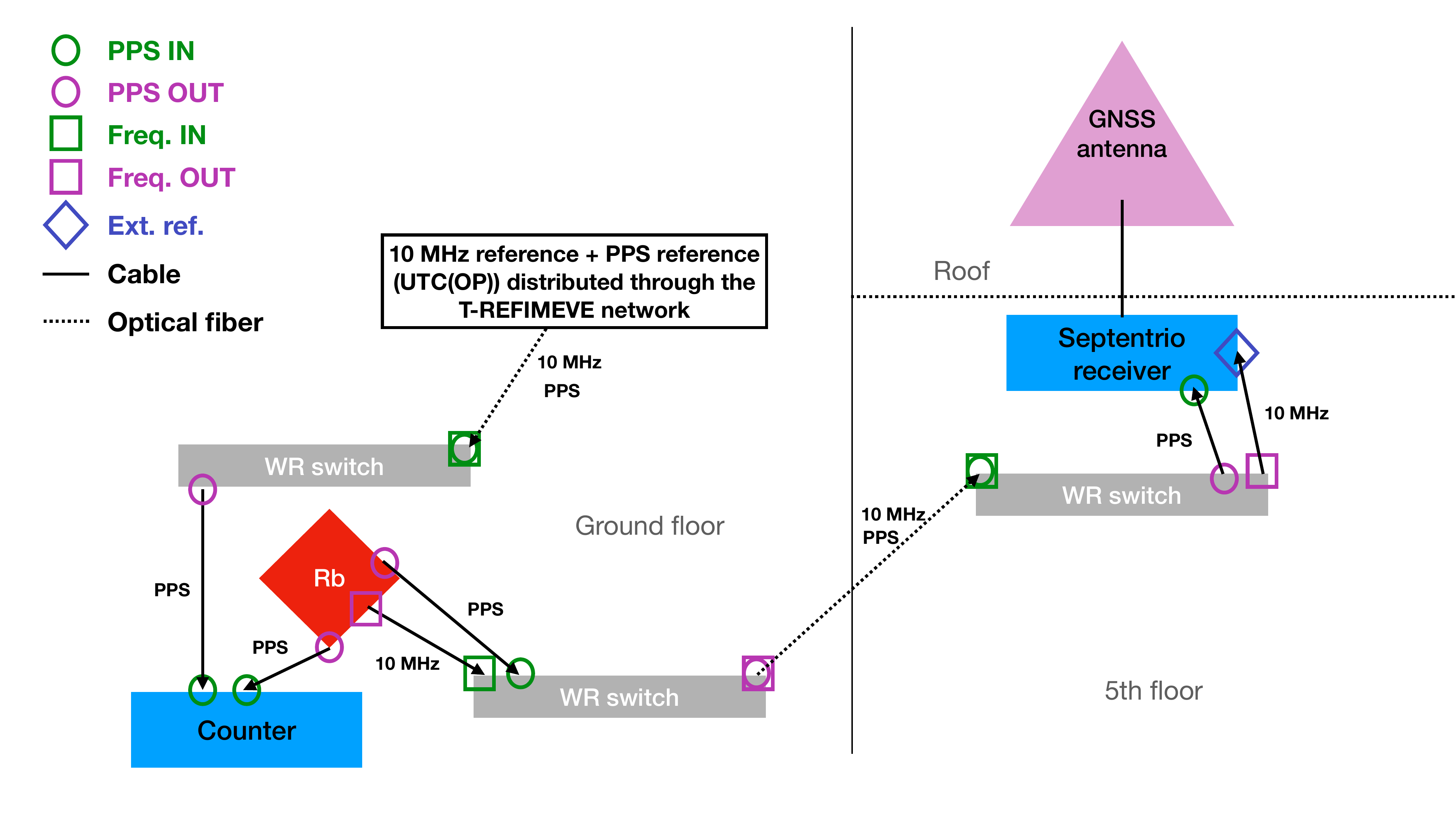}
\centering
\caption{Experimental setup used in this work. Part of the equipment is installed at the ground floor and the other part at the fifth floor. The relevant signals generated at the ground floor are transported to the fifth floor via optical fibers with the White Rabbit (WR) protocol. This particular setup mimics what could happen in underground experiments where the clock signal would be generated underground whereas the GNSS antenna and receiver would be located above-ground. \label{fig:setup}}
\end{figure}

The experimental setup that we used is schematized in Figure~\ref{fig:setup}.
It is located at the Pierre and Marie Curie (Jussieu) campus of the Sorbonne University in Paris. The setup consists of two main parts: one represents the timing generation and correction setup, that could be reproduced in the HK experiment, and the second part is related to testing the efficiency of the correction method. In the first part a Rubidium clock (Rb) in free-running mode, at the ground floor of the laboratory, generates a Pulse Per Second (PPS) signal and a $10$~MHz signal that are transported to the fifth floor with the White Rabbit (WR) protocol~\cite{WR}. The timing signals of the slave WR switch are used by a GNSS receiver as a reference for its internal clock. The receiver connected to its antenna on the roof, above the fifth floor, is used to measure time comparisons between GPS Time and the Rubidium clock. This physical distance between the time generation part and the receiver was done on purpose to mimic what would happen in many long-baseline physics experiments. Indeed, in Hyper-Kamiokande, the Rubidium clock would be placed inside a mountain, where a cavern has been dug to host the detector, whereas the receiver would have to be placed outside in a valley. The second part of our experimental setup is contained in the experimental room at the ground floor and its purpose is to validate the performance of the method and would thus not be reproduced in the final setup in Hyper-Kamiokande. It consists of a counter measuring the time difference between the Rubidium clock PPS signal and the French realization of UTC (called UTC(OP) for Observatoire de Paris). The UTC(OP) reference signals are available at the laboratory that is connected to the T-REFIMEVE network~\cite{Refimeve,Refimeve2} via a third WR switch.

\subsubsection{Rubidium clock}

The Rubidium atomic clock used is the FS725 Rubidium Frequency Standard sold by \href{https://www.thinksrs.com/products/fs725.html}{Stanford Research Systems} integrating a rubidium oscillator of the PRS10 model. It provides two $10$~MHz and one $5$~MHz signals with low phase white noise and its stability estimated via the Allan Standard Deviation (ASD) \cite{OASD} at $1$~s is about $2\times10^{-11}$. It also provides a PPS output with a jitter of less than $1$~ns. Its $20$ years aging was estimated to less than $5\times 10^{-9}$ and the Mean Time Before Failure is over $200,000$ hours. In this work we use the Rubidium clock in free-running mode but it can also be frequency disciplined using an external 1 PPS reference, based on GPS for instance. The FS725 is installed at the ground floor of our laboratory and its $10$~MHz and $1$ PPS outputs are transported to the GNSS receiver at the fifth floor.

\subsubsection{White Rabbit switches}
The White Rabbit (WR) project  \cite{WR} is a collaborative effort involving CERN, the GSI Helmholtz Centre for Heavy Ion Research, and other partners from academia and industry. Its primary objective is to develop a highly deterministic Ethernet-based network capable of achieving sub-nanosecond accuracy in time transfer. Initially, this network was implemented for distributing timing signals for control and data acquisition purposes at CERN's accelerator sites.
The described experimental setup uses two WR switches to propagate with great precision the Rubidium clock PPS and frequency signals from the ground floor to the fifth floor.

\begin{figure}[t]
\begin{center}
\includegraphics[width=\textwidth]{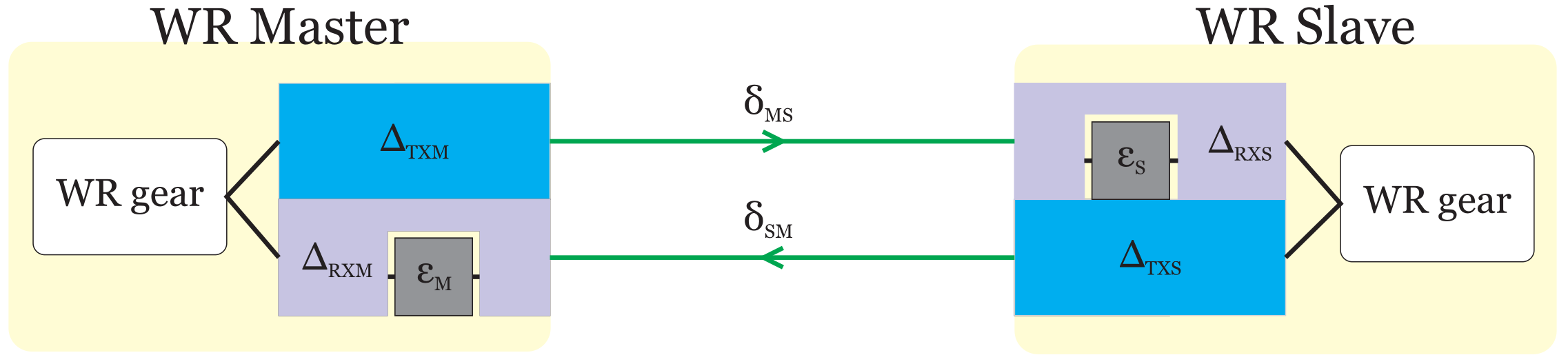}
\end{center}
\caption{White Rabbit link model, from \cite{WRCalibration}}
\label{fig:wrlinkmodel}
\end{figure}

\begin{figure}
\begin{center}
\includegraphics[width=0.8\textwidth]{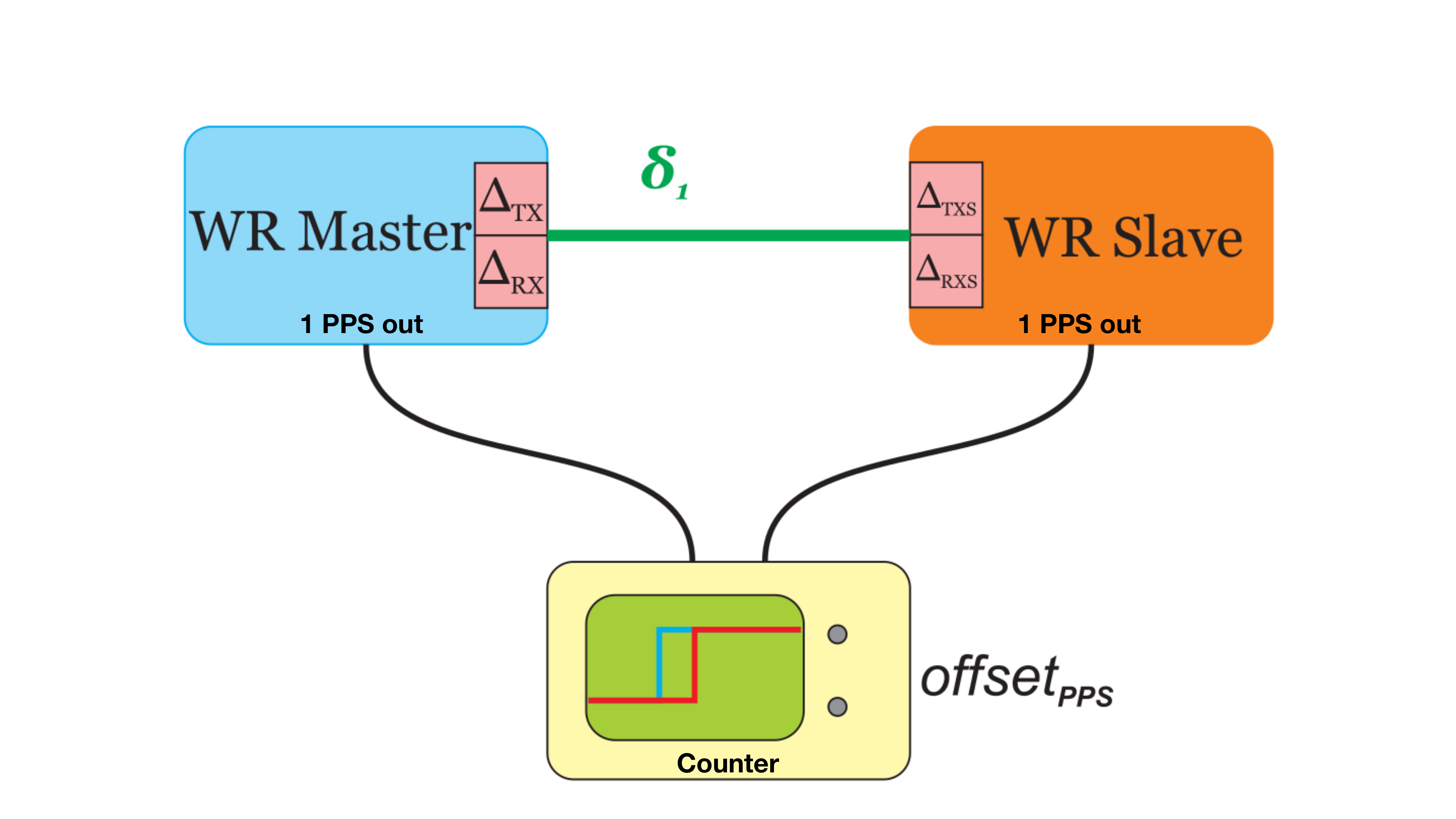}
\includegraphics[width=\textwidth]{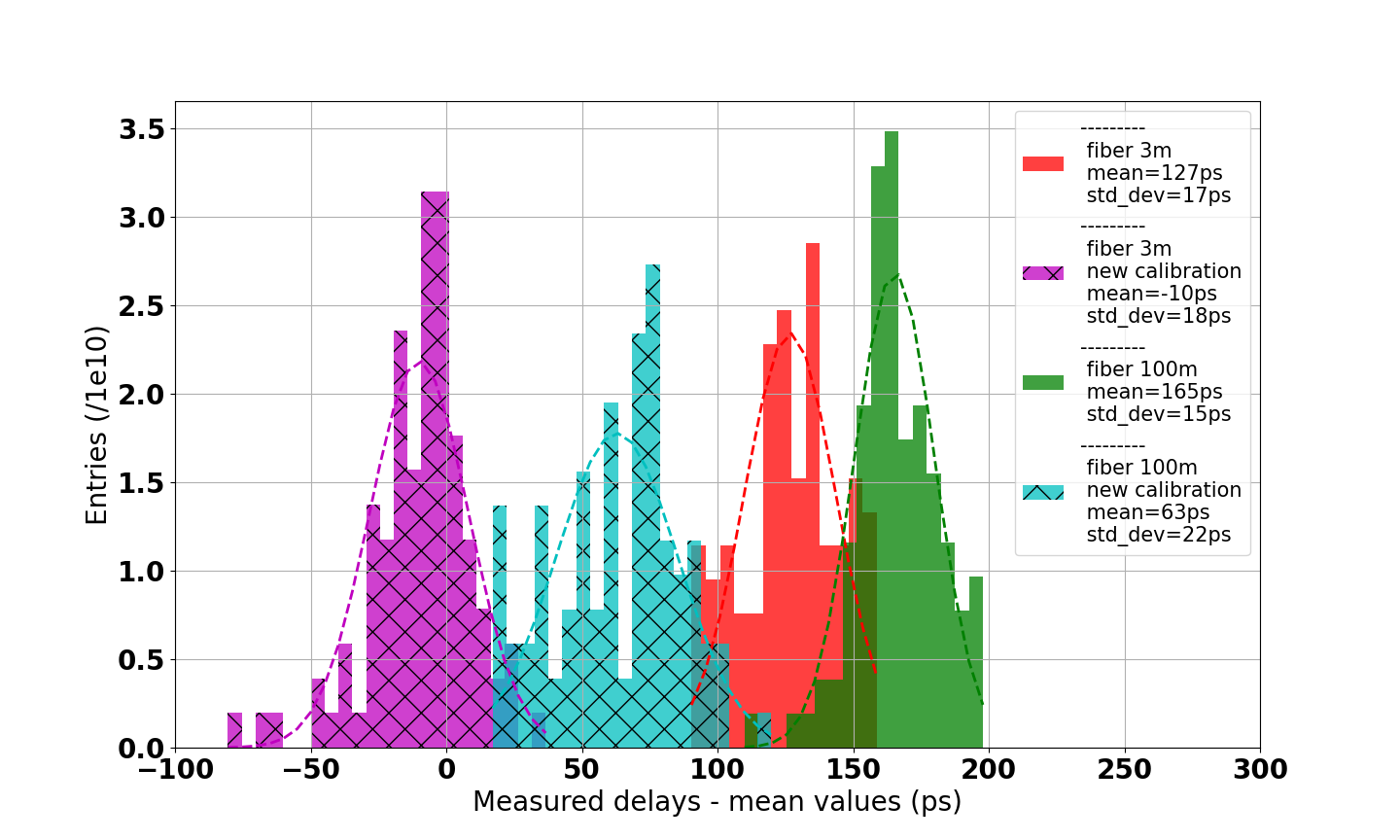}
\end{center}
\caption{Difference between the PPS OUT signals of the White Rabbit slave and master switches before and after calibration}
\label{fig:wrcalibration}
\end{figure}

The calibration of the link allows to obtain a sub-nanosecond synchronization between switches. A White Rabbit link between two devices is characterized by specific hardware delays and fiber propagation latencies.
Each WR Master and WR Slave possesses fixed transmission and reception delays ($\Delta_{TXM}$, $\Delta_{RXM}$, $\Delta_{TXS}$, $\Delta_{RXS}$). These delays are the cumulative result of various factors such as SFP transceiver, PCB trace, electronic component delays, and internal FPGA chip delays. Additionally, there is a reception delay on both ends caused by aligning the recovered clock signal to the inter-symbol boundaries of the data stream, referred to as the bitslide value ($\epsilon_M$ and $\epsilon_S$ in Figure \ref{fig:wrlinkmodel}).
We can see the results of calibration process using a counter in Figure \ref{fig:wrcalibration}, the difference of PPS signals between the WR slave and master switches changes from 165~ps to 60~ps (with a 100~m long fiber). Delays introduced by the cables were subtracted to the mean values. 

As a part of the \href{https://www.refimeve.fr/index.php/fr/presentation/com-allevents-settings/t-refimeve.html}{T-REFIMEVE} network~\cite{Refimeve,Refimeve2}, the LPNHE has access through a dedicated switch to the official French realization of the UTC, called UTC(OP) (for Observatoire de Paris)~\cite{UTC_OP}, transported from the SYRTE laboratory via White Rabbit protocol. REFIMEVE is a French national research infrastructure aiming at the dissemination of highly accurate and stable time and frequency references to more than 30 research laboratories and research infrastructures all over France. The reference signals originate from LNE-SYRTE and are mainly transported over the optical fiber backbone of \href{https://www.renater.fr/en/accueil-english/}{RENATER}, the French National Research and Education Network.

\subsubsection{Counter}

The counter is the \href{https://www.keysight.com/us/en/assets/7018-02642/data-sheets/5990-6283.pdf}{53220A} model from Keysight Technologies. Here it was used to measure the time interval between the two PPS signals: the UTC(OP) PPS reference and the one generated by the free-running Rubidium clock.

\subsubsection{Septentrio GNSS antenna and receiver}

\begin{figure}[tbp]
\begin{center}
\includegraphics[width=0.7\textwidth,keepaspectratio]{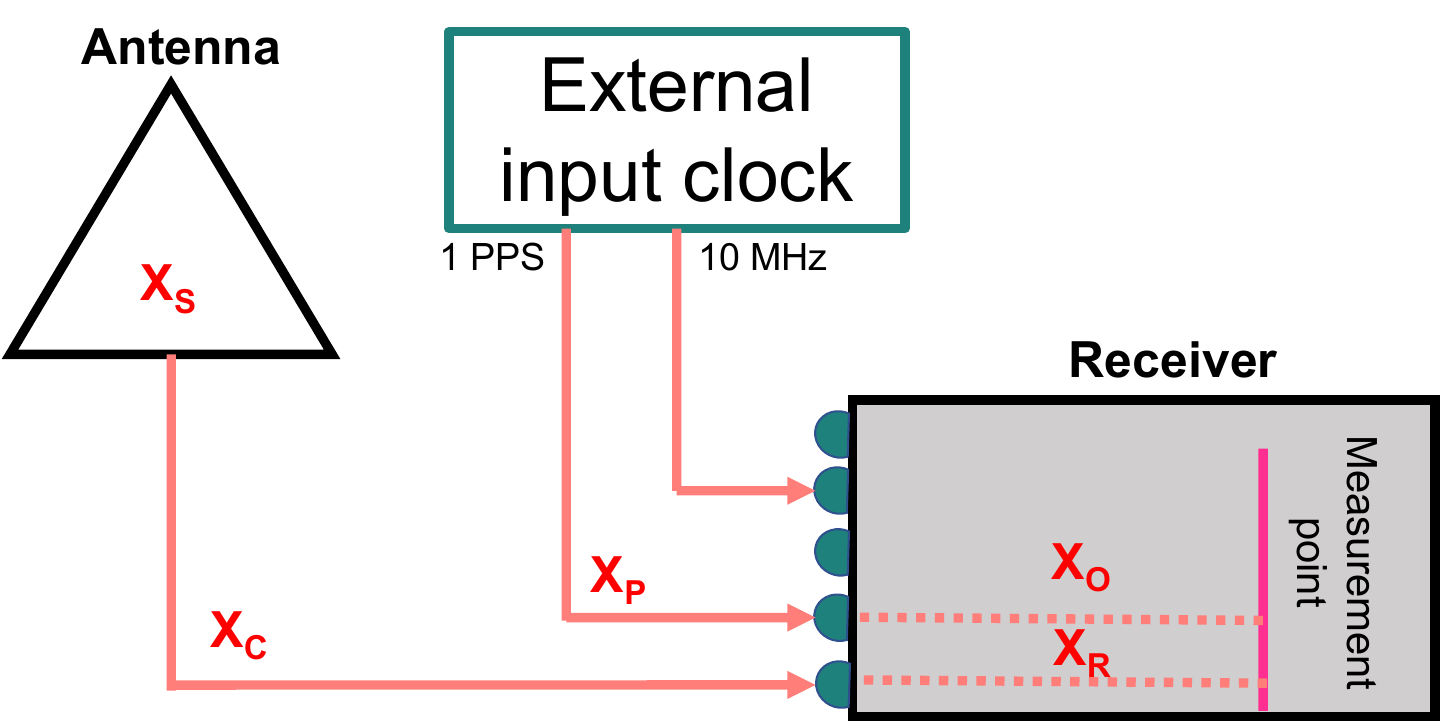}
\end{center}
\caption{Delays to consider for the selected GNSS receiver+antenna pair, from \cite{ThesisMellet}}
\label{fig:delays}
\end{figure}

We use the \href{https://www.septentrio.com/en/products/antennas/polant-chokering}{Septentrio PolaNt Choke ring GNSS antenna} that supports GNSS signals from many satellite constellations including GPS, GLONASS, Galileo and BeiDou. In this work, we restrict the analysis to GPS but it can easily be generalized to any subset of constellations. The antenna position has been previously measured to a precision better than 6~mm by trilateration with the help of a web-based service provided by the Canadian government~\cite{nrcan}. 
We use a \href{https://www.septentrio.com/en/products/gnss-reference-receivers}{Septentrio PolaRx5 GNSS reference receiver} as a timing receiver to compare GPS Time to the Rubidium clock. The receiver performs measurements based on the $10$~MHz reference signal coming via White Rabbit from the Rubidium clock. The Rubidium clock $1$ PPS signal, also transported to the receiver via White Rabbit, is used at initialization to identify the $10$~MHz cycle. Note that this $1$ PPS input is kept during the whole data-taking to avoid possible phase jumps due to perturbations. The Septentrio receiver provides one measurement every $16$~min which is the middle point of the linear function fitted from the $13$~min of data from the beginning of this $16$~min time window. The results of the measurements are registered using the CGGTTS file format~\cite{CGGTTS}.\\ 
Before taking measurements, the whole system has been calibrated against official reference signals from the SYRTE laboratory. As it can be seen in Figure~\ref{fig:delays}, the following delays need to be measured and taken into account during operation~\cite{DelaysReceiver}. The calibration procedure~\cite{Link_calib} consists in measuring these: 
\begin{itemize}
  \item $\mathrm{X_S}$: internal delay inside the antenna, frequency dependent
  \item $\mathrm{X_C}$: delay caused by the antenna cable 
  \item $\mathrm{X_R}$: internal delay of the receiver for the antenna signal, frequency dependent
  \item $\mathrm{X_P}$: in case an external signal is given in input, connection cable delay
  \item $\mathrm{X_O}$: in case an external signal is given in input, internal receiver delay between external 1~PPS and internal clock
\end{itemize}

$\mathrm{X_S}$ and $\mathrm{X_R}$ depend on the GNSS carrier frequency that is being tracked, meaning it is specific to each frequency of each GNSS constellation. The calibration was performed for both GPS and Galileo constellations, each having two available carrier frequencies.  
The cable delays $\mathrm{X_C}$ and $\mathrm{X_P}$ were evaluated with an oscilloscope by sending a pulse in the cable and measuring the timing of the reflection.
To reproduce the experimental conditions of underground experiments like HK or DUNE where the GPS antenna is outside, away from the detector, a $100$~m cable was used and calibrated. The total cable delay was measured to be $505$~ns. The internal delays of the antenna and receiver can only be measured together (for each frequency) as $\mathrm{INTDLY = X_S + X_R}$. This was done through a comparison with OP73, one of the calibrated GNSS stations of SYRTE, and with UTC(OP), the French realization of UTC, as an input to the two receivers. The values of INTDLY found for the two most widely available carrier frequencies of the GPS constellation (L1 and L2) and the Galileo constellation (E1 and E5a) are given in Table~\ref{tab:INTDLY}.\\
\begin{table} [h]
\begin{center}
\caption{Values of INTDLY in ns found for the first antenna+receiver system calibrated at the SYRTE laboratory against the OP73 station}
\begin{tabular}{|c|c|c|c|}
\hline
GPS L1 & GPS L2 & Galileo E1 & Galileo E5a  \\
\hline
25.832  & 22.871 & 28.242 & 25.431 \\
\hline \end{tabular} 
\label{tab:INTDLY}
\end{center}
\end{table}

The delays $\mathrm{X_C}$, INTDLY, and REFDLY can then be given as parameters of the receiver so that they are automatically handled in any further use of the receiver. Uncertainties on the measured delays were evaluated to 4 ns according to estimations fixed for the employed method. The calibration needs to be re-done for any new antenna+receiver+antenna cable combination.

\subsection{Corrections methods}
\subsubsection{General principle}
To synchronize the Rubidium time stamps to UTC, we apply a time-dependent correction (quadratic or linear) to the time series generated by the free-running Rubidium clock $\phi_{Rb}(t)$. We model the $k^\text{th}$ portion of the time series ($dt_{Rb,GPS}$), defined as the difference between the free-running Rubidium clock and GPS Time, as a (one or two degrees) polynomial of time\begin{equation}
    \forall t\in[t_{k-1}, t_k],~ dt_{Rb,GPS}(t)=a_k\cdot t^2+b_k\cdot t +c_k.
\end{equation} 
The coefficients $a_k$ ($a_k=0$ in case of linear fit), $b_k$ and $c_k$ of the polynomials are extracted from least square polynomial fits of the time difference distributions. The fits of these differences, obtained from the Septentrio receiver, are performed for every $k^\text{th}$ time window of length $\Delta t$. In other words, we model the Septentrio measurements with a piece-wise polynomial function of time. For the $k^\text{th}$ time window (between $t_k$ and $t_{k+1}$), we get the corrected time stamps
\begin{linenomath}
\begin{equation}
    \forall t\in [t_k, t_{k+1}],~ \phi_{Rb,corr}(t) = \phi_{Rb}(t)-a_k\times t^2-b_k\times t -c_k .
\end{equation}
\end{linenomath}
 The time-length $\Delta t$ of the pieces (time windows) has to be chosen carefully. In particular, it should be short enough in order to correct for the effect of the frequency random walk of the Rubidium clock.

\begin{figure}[tbp]
    \centering
    \includegraphics[width=\textwidth]{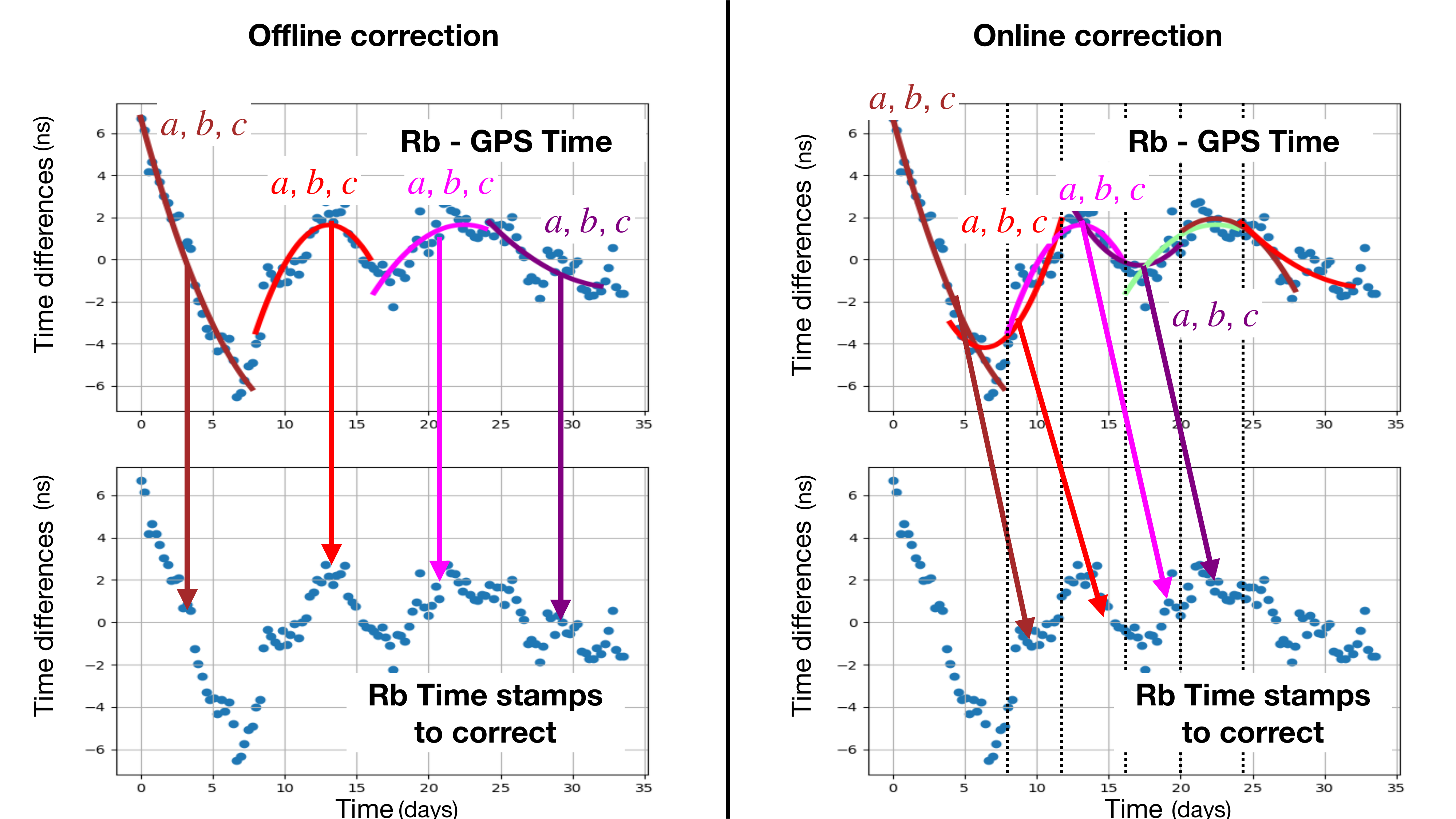}
    \caption{Schematic representation of the offline (left) and online (right) corrections. In the offline correction, we extract the correction coefficients using Rubidium - GPS Time comparison from the same time-window as the data we want to correct. In the online correction,  we use Rubidium - GPS Time comparison from the previous time-window with respect to the data interval we want to correct. Only the second correction can be applied in real time as it only requires comparisons with GPS Time from previous measurements.}
    \label{fig:offlineVSonline}
\end{figure}

In the following, we consider two types of correction: the offline and the online corrections. The difference between the two methods is illustrated in Figure~\ref{fig:offlineVSonline}. 
The offline correction consists in using the Septentrio data from the same time-window as the Rubidium signal to extract the $a_k$, $b_k$ and $c_k$ coefficients. This correction is called offline because it requires the Septentrio data from up to $t_k+\Delta t=t_{k+1}$ to correct all the time stamps between $t_k$ and $t_{k+1}$ so it cannot be performed in real-time (one would need to wait a time $\Delta t$ to extract the correction coefficients for the $t_k$ time stamp).

The online correction consists in correcting the Rubidium time stamps between $t_k$ and $t_{k+1}$ using Septentrio data collected before $t_k$. One example of online correction is illustrated in Figure~\ref{fig:offlineVSonline} where overlapping windows are used.
\begin{figure}[tbp]
    \centering
    \includegraphics[width=\textwidth]{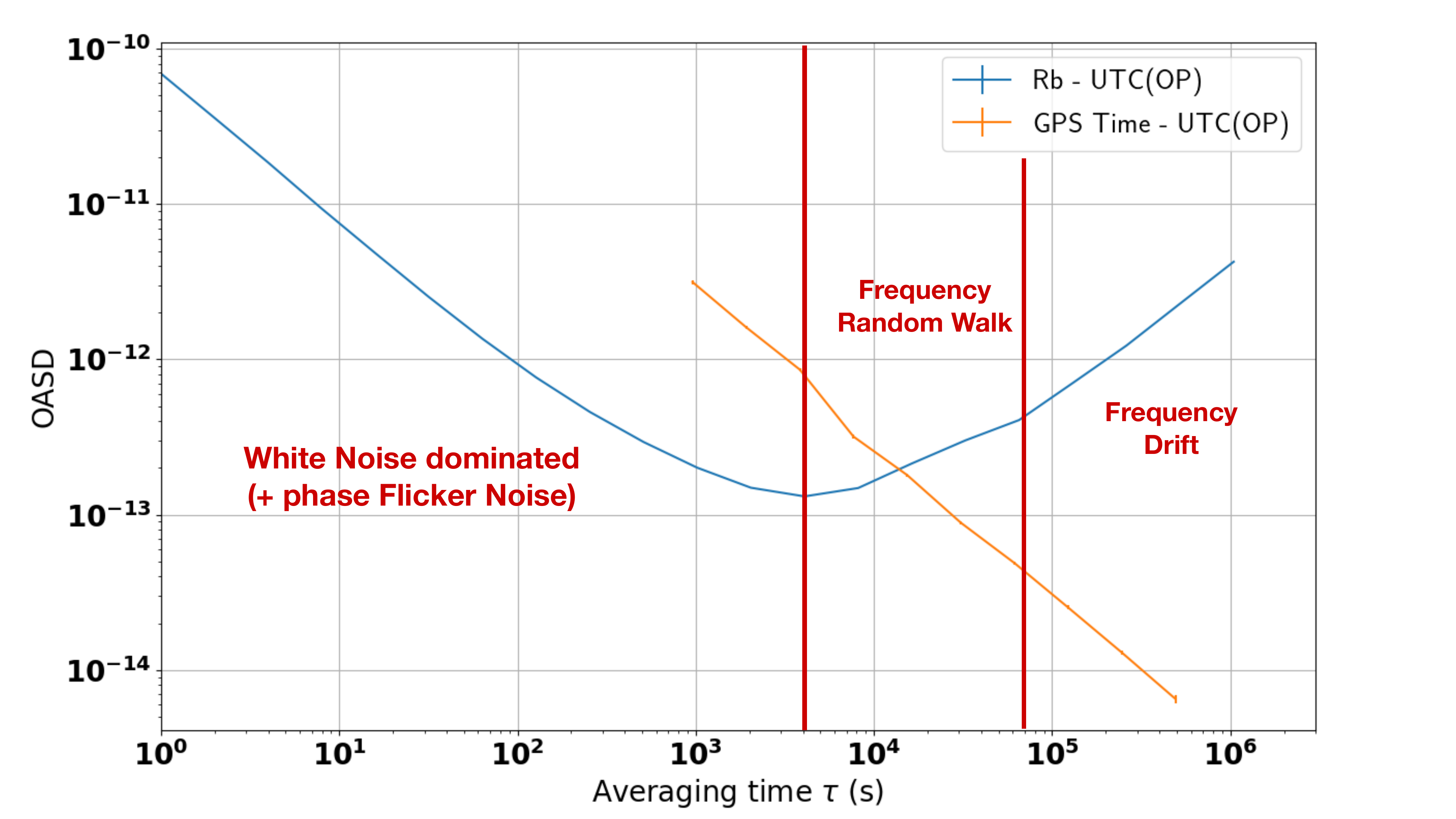}
    \caption{Overlapping Allan Standard Deviation of the Rb vs UTC(OP) time difference (in blue), measured by the counter, before any correction, and of GPS Time vs UTC(OP) (in orange) measured by the Septentrio receiver. The main types of noises affecting the Rubidium clock stability are indicated where they are limiting the stability.}
    \label{fig:ASD_char_Rb}
\end{figure} 
This method is called online because it can be applied in real time. In the following, we will consider the most frequent possible update of the $a_k$, $b_k$ and $c_k$ coefficients: they will be updated every time we receive a new data point from the Septentrio receiver (every $\delta t \approx 16$ minutes in our case). This means that we have $t_{k+1}=t_{k}+\delta t$ so that the $a_k$, $b_k$ and $c_k$ coefficients are extracted using Septentrio data between $t_k-\Delta t$ and $t_k$ and are used to correct the time stamps between $t_k$ and $t_k+\delta t$. In that particular case every Septentrio data point will have been used in multiple fits, the number depending on the length of the fit time window $\Delta t$. 

The performance of the correction is evaluated in two ways. First, we look at the stability of the corrected time series estimated with the Overlapping Allan Standard Deviation (OASD). Then, we also look at the time difference against GPS signal after correction.

\subsubsection{Validation of the method with simulations}
Before evaluating the performance of our timing system when integrating the correction algorithm, the method was validated on simulated signals \cite{ThesisMellet} in order to isolate the effect and performance of the correction from any measurement issues.

\paragraph{\textbf{Simulation details}}

Two types of signal comparisons to a perfect reference were simulated: a free running Rubidium clock and a GPS time signal, as measured by the Septentrio receiver. The quadratic time drift due to the Rubidium clock frequency drift was not included because it is deterministic and is therefore not challenging to correct. At first order, the clock signal can be modeled by white noise ($WN$) in both phase and frequency as well as a random walk ($RW$) noise in frequency. Based on the characterization of the Rubidium clock, the phase and frequency flicker noises can be neglected for this purpose. Indeed, the characterization of our Rubidium clock in Figure \ref{fig:ASD_char_Rb} showed that the frequency flicker noise had a negligible impact on the OASD. Furthermore, the phase white and flicker noises have a similar impact on the standard OASD and cannot be distinguished here. We chose to ignore the phase flicker noise as it is less straightforward to simulate and it should not impact the long term random walk drift that we want to correct. GPS Time can be modeled as pure phase white noise. The corresponding OASD as a function of the averaging time $\tau$ can be modeled \cite{ASDNoise1,ASDNoise2,ASDNoise3} by:
\begin{equation}
\label{eq:OASDNoise}
OASD(\tau) \cong  A_{WNp} \times \tau^{-1} + A_{WNf} \times \tau^{-1/2} +A_{RWf} \times \tau^{+1/2}.
\end{equation}
The amplitudes $A$ of these main frequency and phase noises were determined through fitting this model (Eq. \ref{eq:OASDNoise}) to the OASD of the data when characterizing our equipment (see Figure \ref{fig:ASD_char_Rb}) and found to be:
\begin{eqnarray}
A_{WNf}&=& 7\times 10^{-12}~s^{1/2},  \\ \nonumber
A_{RWf}&=& 1\times 10^{-15}~s^{-1/2}, \\
A_{WNp}&=& 5\times 10^{-11}~s, \nonumber
\end{eqnarray}
for the free-running Rubidium clock and for GPS Time:
\begin{eqnarray}
A_{WNf}&=& 0~s^{1/2}, \\ \nonumber
A_{RWf}&=& 0~s^{-1/2}, \\ 
A_{WNp}&=& 2\times 10^{-9}~s, \nonumber
\end{eqnarray}
with indices $f$ and $p$ for frequency and phase respectively.
Using random numbers generation and a model with these types of noise discussed just above, time series were simulated.

The equivalent of $10^6$ s of data was simulated. 
To mimic the output of the GNSS receiver, time differences between the simulated Rubidium clock and the simulated GPS Time ($\Delta t^{i}_{Rb-ref}$) are computed every 16~mn.

\paragraph{\textbf{Offline corrections}}

First, the offline corrections were tested on the simulated data. 
\begin{figure}[tbp]
\begin{center}
\includegraphics[width=10cm,keepaspectratio]{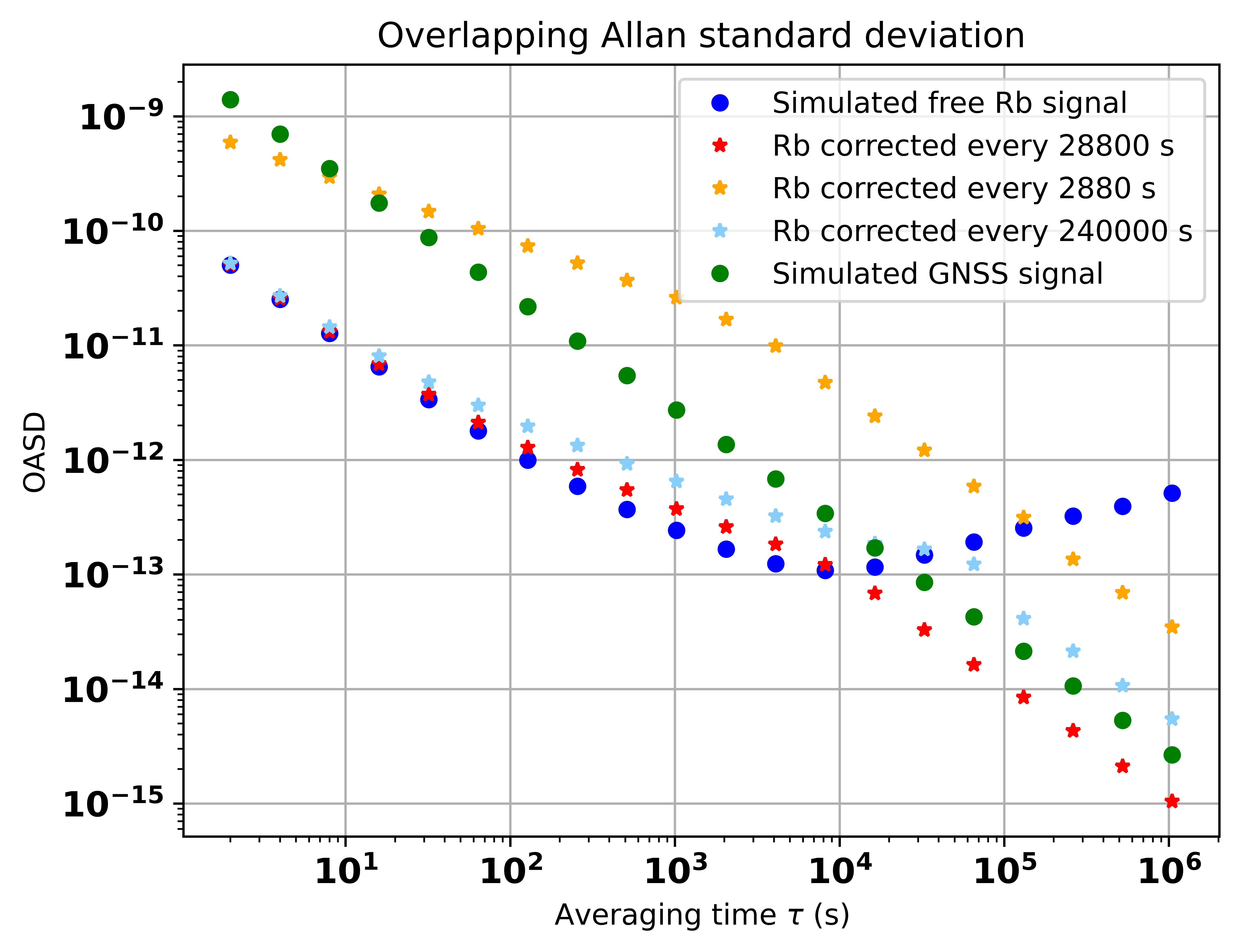}
\end{center}
\caption{Comparison of overlapping ASD for corrected signals, with offline correction, with different time windows}
\label{fig:OfflinecorrDiffT}
\end{figure} 
In Figure~\ref{fig:OfflinecorrDiffT}, the uncorrected simulated signals of the GPS and the clock are reported in dotted symbols for comparison. The increase of the clock's OASD after $\tau = 10^4$~s due to the random walk is clearly visible. Looking at the OASD of the corrected signals (starred symbols), one can see that the random walk is eliminated at longer terms  which indicates a success of the correction method (quadratic). Moreover, one can determine that the ideal length $\Delta t$ of the correction time windows lies around $3\times10^4$~s which corresponds logically to the intersection of the free-running Rubidium clock and GPS Time OASD curves. Indeed, the red curve with a time window of 28800~s shows an ideal combination of the short-term stability of the clock and the absence of random walk at longer scales. On the opposite, the yellow (shorter time window) and light blue (longer time window) curves show respectively a degradation of the short term performance and a remaining random walk component in the region between $\tau = 10^4$~s and the time window length (here 240000~s). Note that since we use the GNSS signal to correct the Rubidium signal, one would expect the corrected signal's OASD at larger $\tau$ to be limited by the GNSS signal stability. However, since we here model the behavior of the time difference between the Rubidium and the GNSS Time with a piece-wize polynomial function of time, the fitted behavior can be smoother (less white noise) than the real distribution. As a result, the corrected signal stability can sometimes be slightly better than the GNSS signal stability, as it is the case when we use the $2880$~s time window (red stars).

\paragraph{\textbf{Online corrections}}

\begin{figure}[tbp]
\begin{center}
\includegraphics[width=10cm,keepaspectratio]{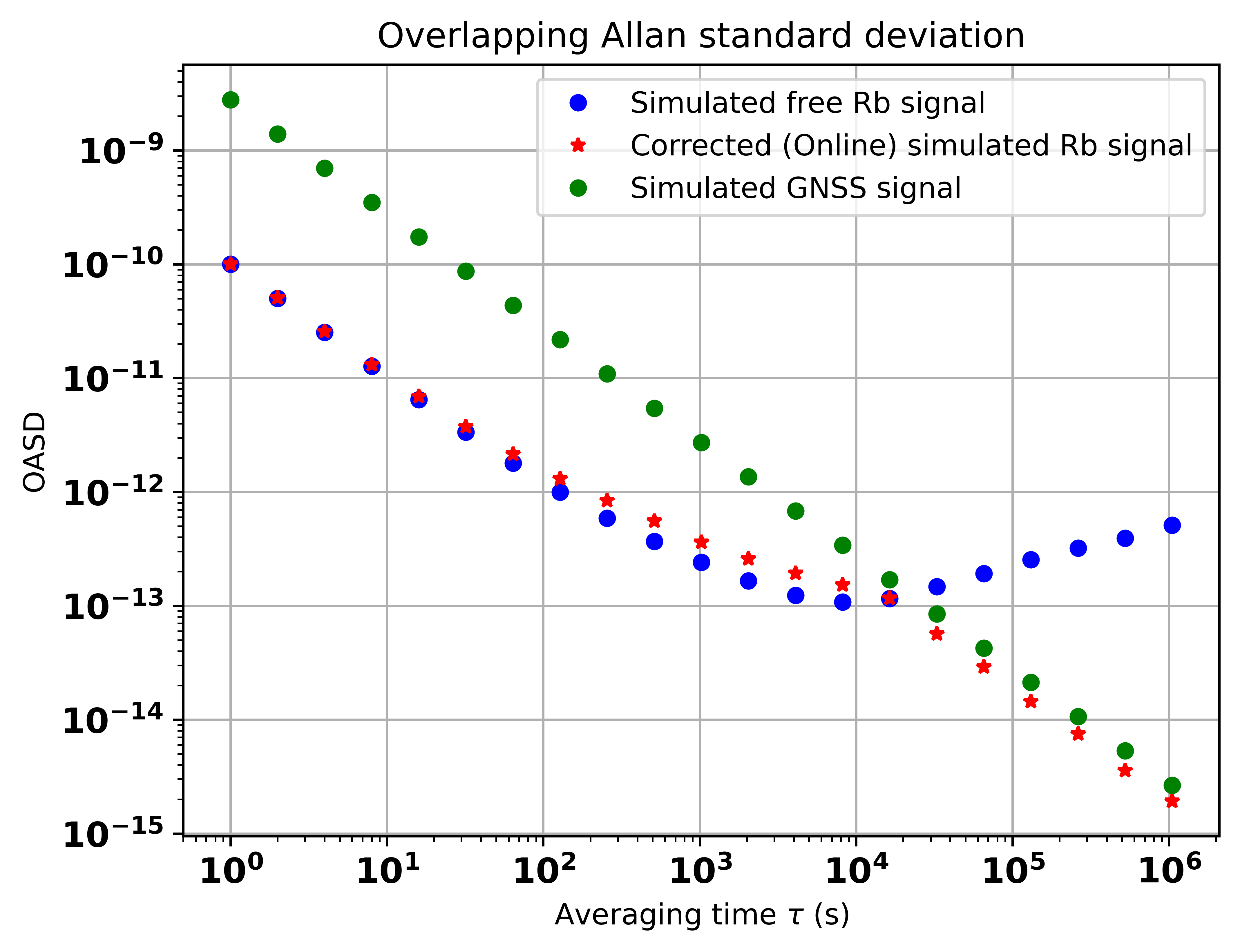}
\end{center}
\caption{After online corrections at $3\times10^4$ s: Overlapping ASD with respect to perfect signal}
\label{fig:Onlinecorr}
\end{figure}

The online (linear) correction method was then applied to the simulated data using time series directly and a correction window length of  $\Delta t = 3\times10^4$~s. The results are shown in Figure~\ref{fig:Onlinecorr} in red and prove to be just as efficient as the offline correction method to remove the random walk at longer time scales which is the main goal. The overall precision on the long term region (after $\approx 10^3$~s) is as expected slightly degraded compared to the offline correction.

\paragraph{\textbf{Conclusion on simulation}}

\begin{figure}[tbp]
\begin{center}
\includegraphics[width=10cm,keepaspectratio]{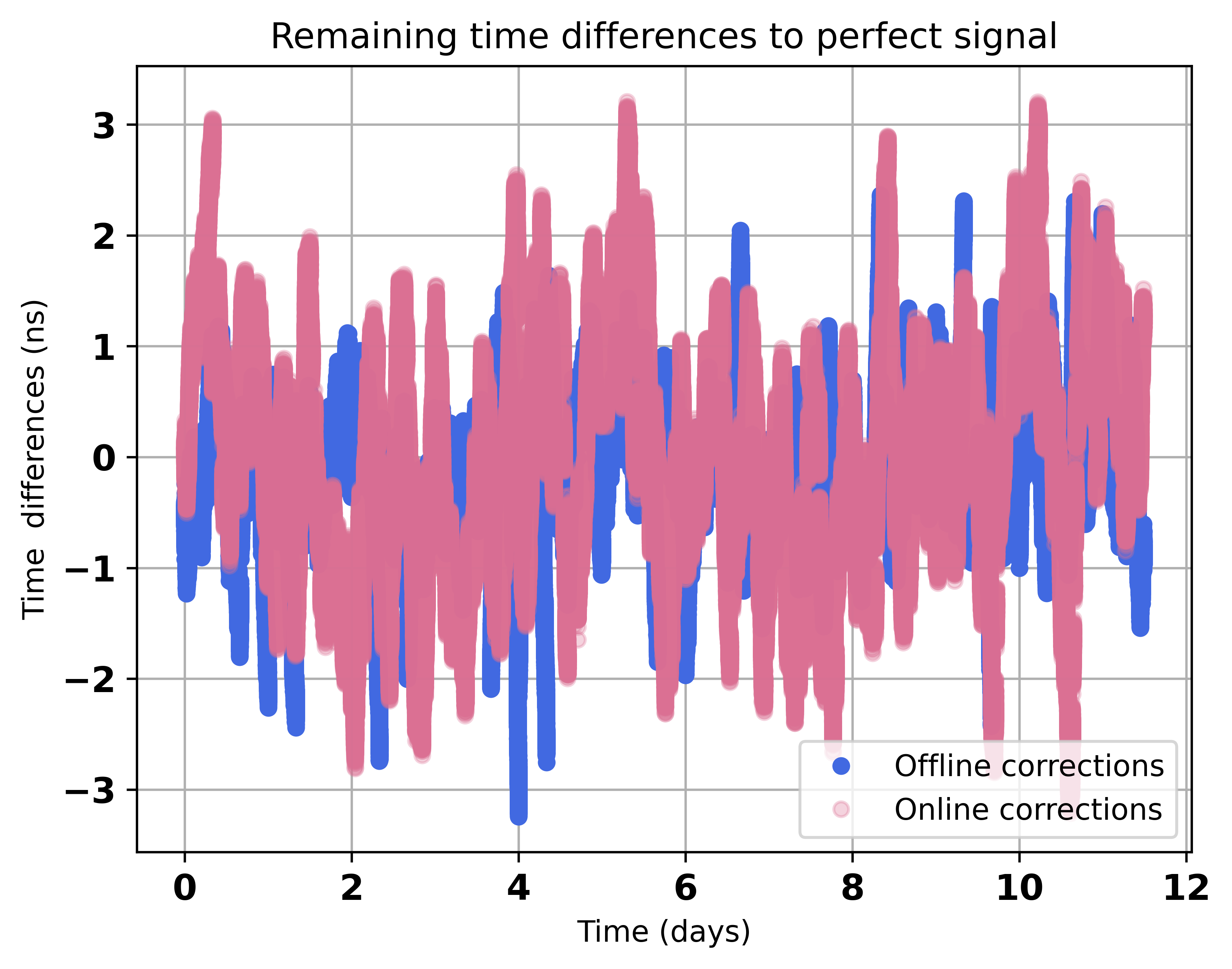}
\end{center}
\caption{Comparison of time variations for simulated signals corrected with the offline method (blue) or with the sliding interval online method (pink)}
\label{fig:OffOncorrcomp}
\end{figure}

As a conclusion, it can be said that the application of the correction algorithms to the simulated signals allowed us to validate the chosen correction methods, both the offline and online ones. Indeed, looking at the residuals after correction in Figure~\ref{fig:OffOncorrcomp}, one can see that the remaining variations for both methods are well within the experimental requirements as they stay within a few ns. Seven different simulations were produced to take into account statistical fluctuations and the remaining time variations were found to be for offline and online corrections respectively $\sigma_{Off} = 0.64 \pm 0.06$~ns and $\sigma_{On} = 1.15 \pm 0.07$~ns.\\
Finally, it is important to note that although this validates the methods for application on data, those are simplified simulations, in particular because only the two noise types are taken into account. As a result, we do expect differences of performance of the correction on real data. It is also possible that the optimal time window for the correction is slightly different for real data because the simulations are not exact representation of data. Two main differences can be noted: the absence of frequency drift and flicker noises in the simulated Rubidium signal and the fact that we assume a perfect signal to compare the Rubidium signal to when evaluating the OASD. Note that the frequency drift induces a quadratic drift of time signals and should therefore be automatically corrected by our correction method.

\subsubsection{Implementation on data}

To check the impact of the correction we compare the Rubidium clock signal before and after correction to UTC(OP) that we receive at the laboratory via the T-REFIMEVE network. The UTC(OP) time signal plays the role of the perfect signal used for the simulations. In the following, we will quantify the stability of the Rubidium signal using the OASD of a time series (according to equation (10) of \cite{oadev_eq}) consisting of time differences between this signal and UTC(OP).  Measuring this time difference frequently, once per second for instance, will allow to also evaluate the very short term stability of the corrected signal which is not possible with the Septentrio measurements that are integrated over $16$ minutes. We use the counter to provide such a measurement every $1$ second approximately. We then perform a simultaneous correction of the Rubidium - GPS Time, as measured by the Septentrio receiver, and of this measured time series. Comparing the OASD of the corrected time series to the uncorrected one, one can quantify the short term stability (below $16$ minutes) after correction while making sure that the random walk was corrected. We can also use this comparison to optimize the value of $\Delta t$ in order to achieve the lowest Allan Standard Deviation possible at all averaging time.

%%%%%%%%%%%%%%%%%%%%%%%%%%%%%%%%%%%%%%%%%%
\section{Results}

In this Section, we present the results of the correction of the Rubidium clock time stamps obtained for simultaneous measurements of around $35$ days with the Septentrio receiver and the counter.  The data used for this analysis are available at the following page~\cite{data}.
The OASD of the time series measured by the counter is shown in Figure~\ref{fig:ASD_char_Rb}. Note that the statistical uncertainty on the estimated OASD, due to the limited number of samples per averaging time, are included as error bars for both curves (Rb and GPS) but they are too small to be visible. Indeed for the Rb vs UTC(OP) OASD, the statistical uncertainty is at the permil level.  Up to an averaging time of around $4\times 10^3$~s, the stability is limited only by the phase white noise and then by the frequency white noise. After that, the OASD first increases as $\tau^{1/2}$ which is characteristic of the frequency random walk. From $\tau\approx7\times10^{4}$~s, the OASD increases proportionally to $\tau$. This is characteristic of a deterministic frequency drift which can be easily characterized and corrected for contrary to the frequency random walk. In comparison, the OASD of the difference between GPS Time and the UTC(OP) reference PPS signal that we receive from LNE-SYRTE, is only limited by a phase white noise at least up to an averaging time of $5\times10^{5}$~s: the OASD keeps decreasing with the averaging time. At low averaging times, the GPS stability is worse than that of the Rb because of this phase white noise: the GPS OASD is of around $3\times10^{-12}$ at $960$~s compared to around $2\times 10^{-13}$ OASD for the Rubidium clock. However, at around $10^4$~s, the stability of the Rb signal becomes worse compared to GPS Time because of the frequency random walk and drift of the Rubidium clock.

\begin{figure}[tbp]
    \centering
        \includegraphics[width=\textwidth]{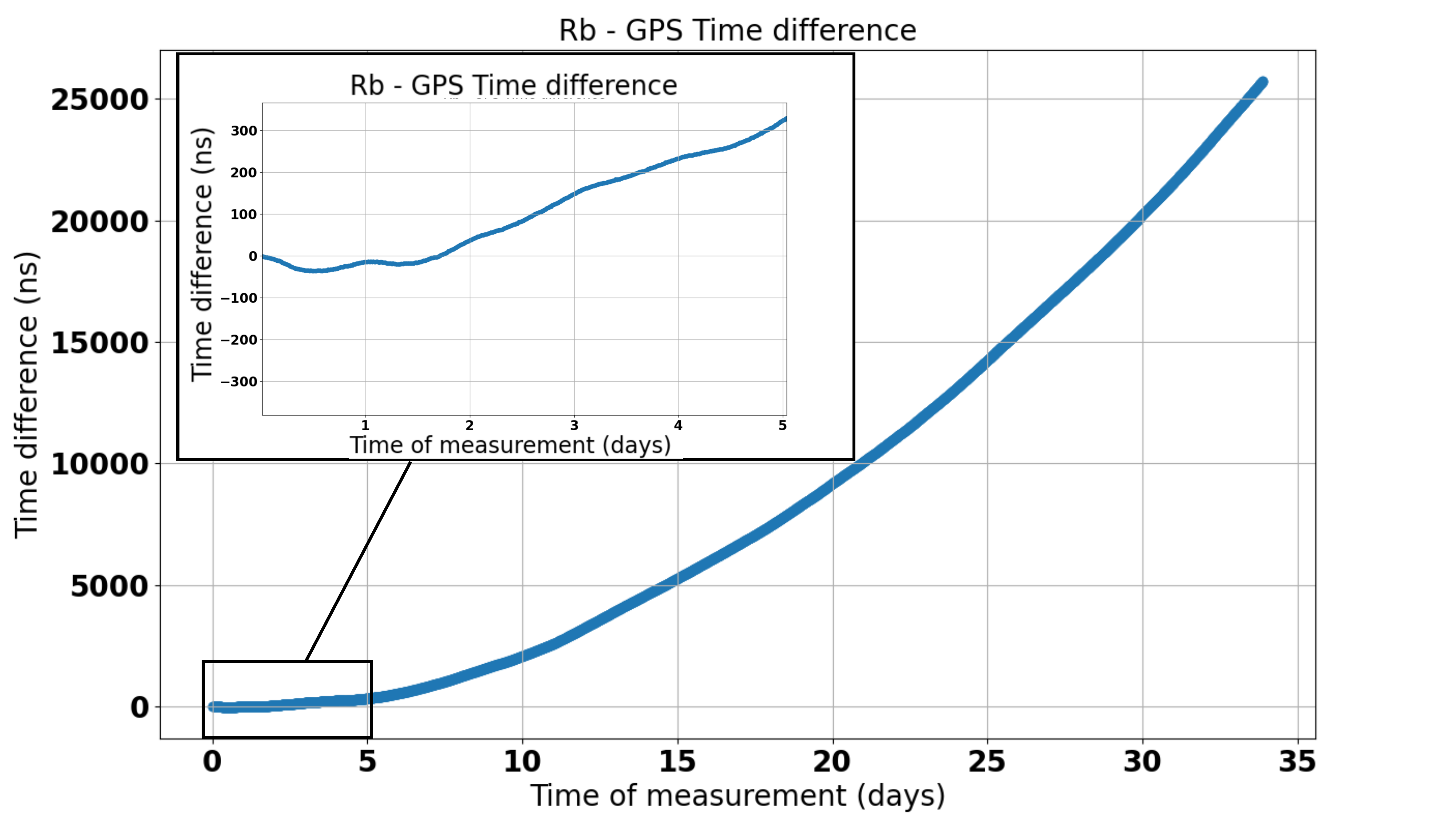}
     \caption{Time difference between the Rubidium clock and GPS Time as measured by the Septentrio receiver. The long term quadratic drift is due to the linear frequency drift of the clock. The zoom on the first five days of data also shows shorter term fluctuations caused by the frequency random walk of the Rubidium clock.} 
    \label{fig:residuals_beforeCorr}
\end{figure}

In this paper, we used only the GPS satellites with an elevation angle (angle between line of sight and horizontal direction) larger than $15^\circ$ to extract the Rubidium time residuals distribution. During the whole data-taking period, for each data point, the Septentrio receiver was able to track an average of $6.5$ GPS satellites and at least $4$ GPS satellites for each data point. To obtain the Rubidium vs GPS Time difference, we take the mean value of the differences between the Rubidium clock and each GPS satellite tracked in the same integration time window of the Septentrio receiver. The obtained time difference is shown in Figure~\ref{fig:residuals_beforeCorr}. 

It shows that the Rubidium clock time signal drifts away from GPS Time in a quadratic function of time because of the frequency linear drift. After around $35$ days, the difference surpasses $25$~$\mu$s. 
A zoom on the first five days of data also shows some shorter term fluctuations characteristic of the frequency random walk. Because of those two sources of frequency variations, we see that after a few days of data-taking, the Rubidium clock time signal can drift away from GPS Time by more than a hundred nanoseconds.

\subsection{Offline correction}

\begin{figure}[tbp]
    \centering
    \includegraphics[width=\textwidth]{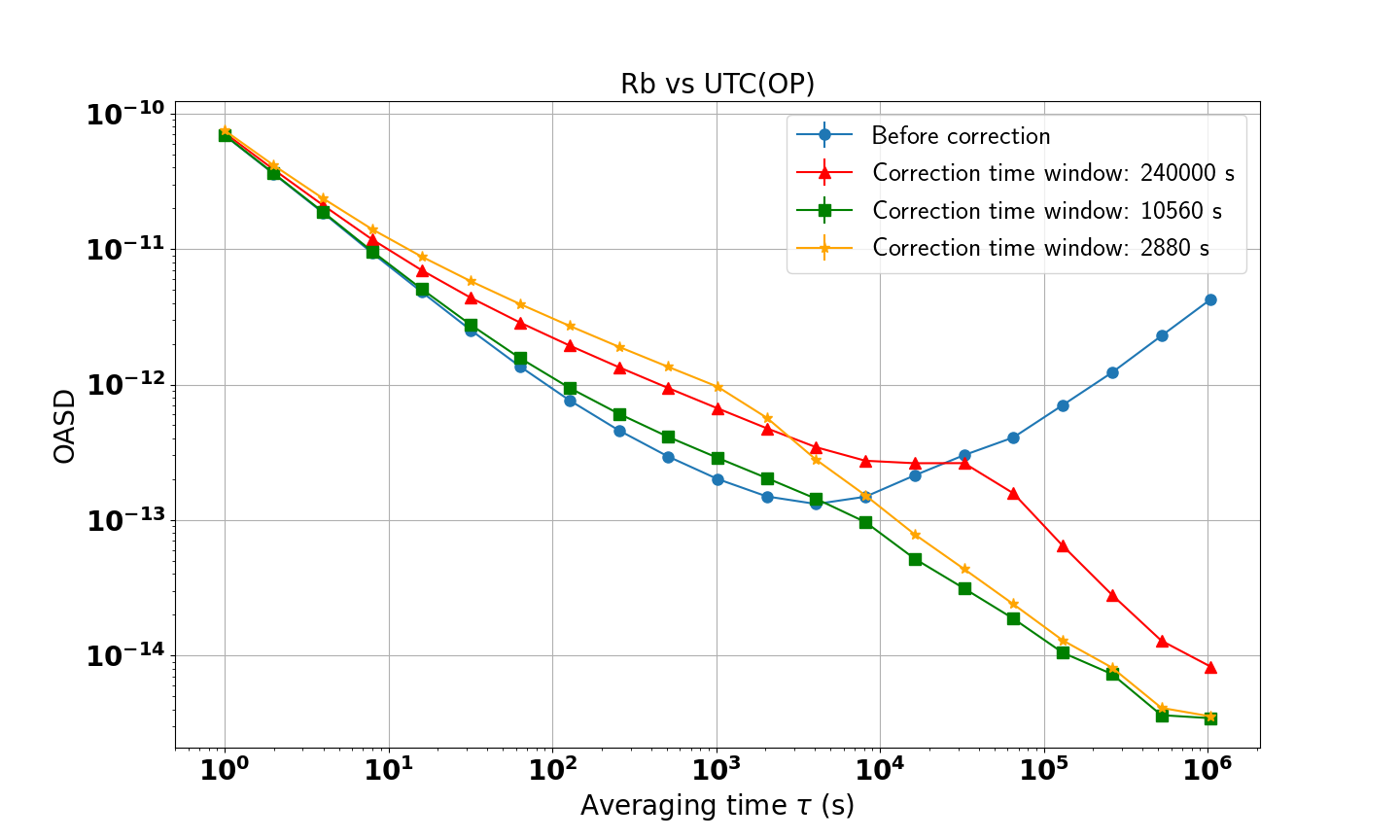}
    \caption{\textbf{Offline Correction}: Overlapping Allan Standard Deviation of the Rb - UTC(OP) time series before correction (in blue) and after the offline correction with a correction time window of $2880$~s (orange), $10560$~s (green) and $240,000$~s (red). The best stability at both short and long averaging times is obtained for the medium time window ($10560$~s$\approx 3$ hours).}
    \label{fig:ASD_offline}
\end{figure}

Figure~\ref{fig:ASD_offline} shows the Allan Standard Deviation of the Rubidium-UTC(OP) data. Note that the measurement rate of the counter was of around $0.995$ measurement per second. The blue curve shows the result for the raw series, before any correction. The other colored curves show the results for the  series corrected offline, with different width of the correction time window. Here, we use quadratic fits of the Septentrio data (so $a_k\neq0$ a priori). The shortest time window ($2880$~s) corresponds to approximately $3$ Septentrio $16$ minutes epochs. The medium ($10560$~s) and largest ($240,000$~s) correspond respectively to $11$ and $250$ Septentrio data points. 

One sees that with the medium time window compared to the two others, we obtain the best stability at all averaging times. At lower averaging times, the performance is very similar to the uncorrected time series. At higher averaging times, the Allan Standard Deviation is much better than the uncorrected series as it keeps decreasing with increasing $\tau$. This is also the case for correction with the shortest time window. This illustrates the fact that both the $2880$~s and $10560$~s windows are able to correct the frequency random walk and linear drift of the uncorrected time series. However, with the shortest correction time window, the short term stability of the time series is degraded compared to the uncorrected series: the value of the ASD at $100$~s increases by a factor $3$. In this scenario, the corrected Rubidium time signal gets very close to GPS Time which is known to have a higher phase White Noise. Finally, the longest correction time window leads to a similar stability as the shortest one for a small $\tau$, and poorer stability at large $\tau$ (above $5\times10^3$~s). 

\begin{figure}[tbp]
    \centering
        \includegraphics[width=\textwidth]{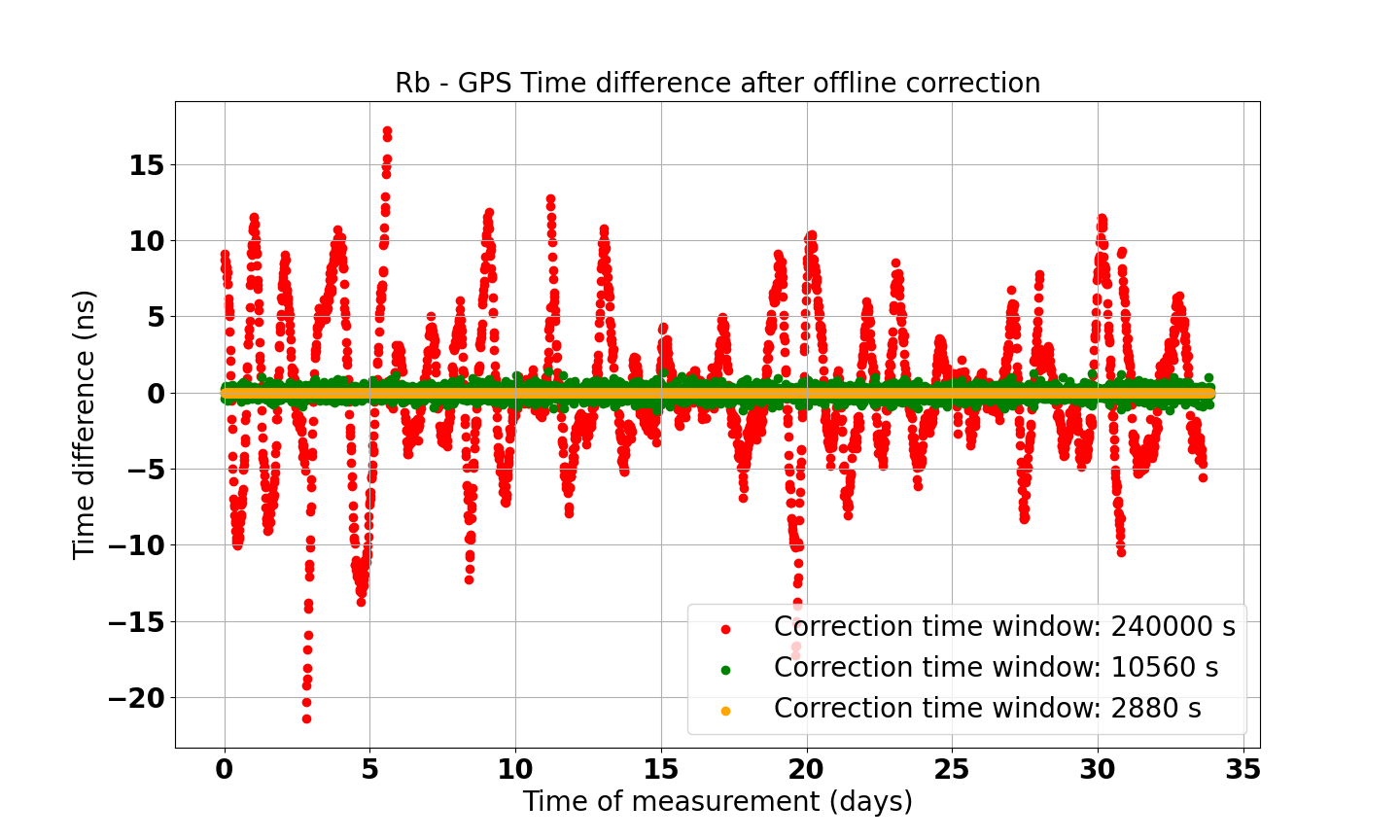}
    \caption{Time difference between the Rubidium clock and GPS Time after the offline correction. Three different correction time windows have been tested: $2800$~s (orange), $10560$~s (green) and $240,000$~s (red). These residuals can be compared to the time difference before correction that were shown in Figure \ref{fig:residuals_beforeCorr}.} 
    \label{fig:residuals_offline}
\end{figure}

Figure~\ref{fig:residuals_offline} shows the Rubidium vs GPS Time difference after the offline correction. 
In offline mode, the shorter the correction time window, the lower the residual differences. However, with the medium length time window, we still get time residuals lower than $3$~ns over the whole data-taking period, which is well below the requirements of HK. With the longest correction time window, jumps of a few tens of nanoseconds are introduced in the time residuals. This explains the overall higher ASD: the stability of the signal is limited by those jumps.  
 The time scale of the variations in the data to fit is too small compared to the $240,000$~s time window. In consequence, the fitted tendency from one piece to another is very different, and the fitted piece-wise polynomial is not continuous. It is also interesting, as a cross-check, to have a look at the fluctuations in the time difference between the Rubidium clock and UTC(OP) after correction. This is summarized in the first line of Table \ref{tab:res_wr} that gives the standard deviation of the time series after correction. The deviations with the two shorter correction time windows are indeed very small (below $2$~ns) confirming that this method can be used for synchronization to UTC. 

With the offline version of the corrections, we thus obtain a very good synchronization to GPS Time at the level of a few nanoseconds with the $10560$~s time window. However, this version of the correction cannot be applied in real time. In the following, we show the results for the online version of the correction that can be applied in real time to correct the time stamps of events in physics experiments.  

\subsection{Online correction}

\begin{figure}
    \centering

        \includegraphics[width=\textwidth]{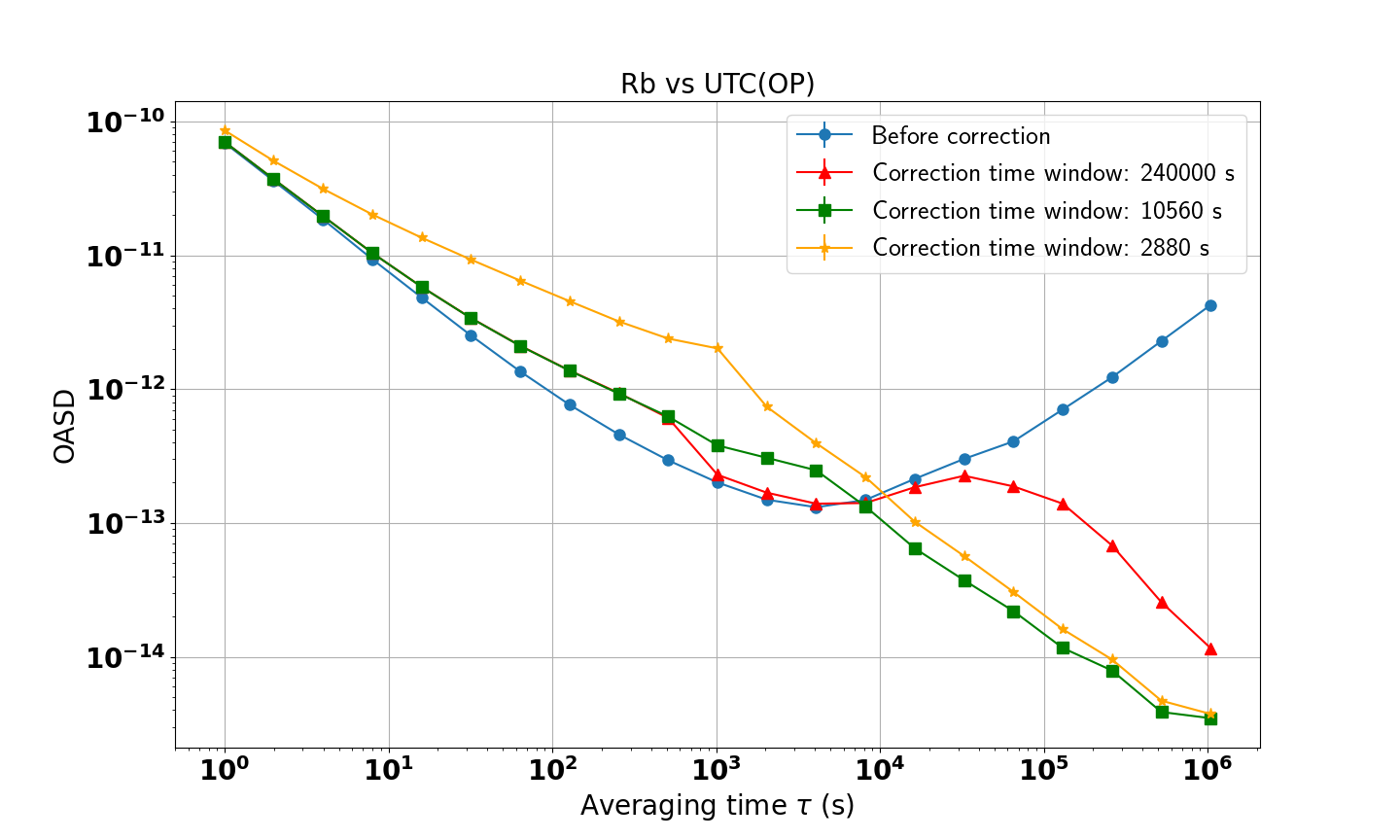}

        \includegraphics[width=\textwidth]{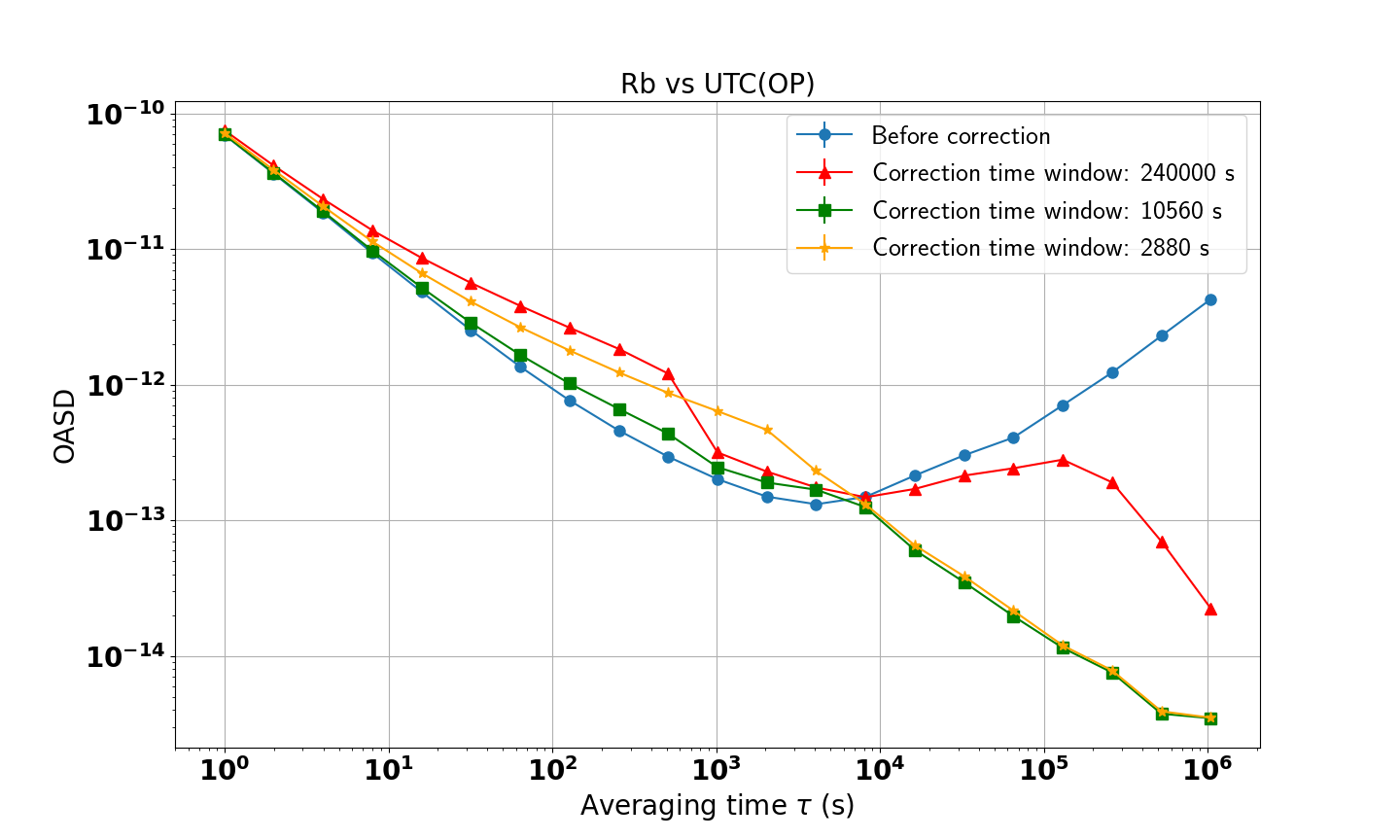}

    \caption{\textbf{Online Correction}: Overlapping Allan Standard Deviation of the Rb - UTC(OP) time series before correction (in blue) and after the online correction with a correction time window of $2880$~s (orange), $10560$~s (green) and $240,000$~s (red). The data were fitted with quadratic (top) or linear (bottom) functions of time. A better stability, similar to the offline correction, can be obtained using linear fits.}
    \label{fig:ASD_online}
\end{figure}

Figure~\ref{fig:ASD_online} shows the Allan Standard Deviation of the uncorrected (blue) and online corrected (other colors) Rubidium - UTC(OP) times series. The same three correction time windows intervals as before are considered. The top panel shows the results using quadratic fits of the Septentrio data and the bottom panel shows the results with linear fits. For the shortest and medium correction time windows, the linear fits lead to better performance with a lower OASD at low averaging times. At $1000$~s, the OASD with the shortest (medium) correction time window is reduced by a factor $2$ to $3$ (resp. $1.5$). 

This behavior can be understood by looking at the number of degrees of freedom (number of data points - number of free parameters) in our fits. For the shortest time windows, the number of degrees of freedom is relatively low ($0$ and $8$) in case of quadratic fits so we risk over-fitting to the past data in order to correct the present data. This number of degrees of freedom is less relevant in the offline correction as the fit is performed on the same data as the correction (the over-fitting is not a problem here). Lowering the number of free parameters is one way of increasing the degrees of freedom hence allowing the fit to better generalize to the present data. Another way to increase the number of degrees of freedom is to increase the number of data points in the fit. For the longest time window, there are $247$ degrees of freedom in the quadratic fit so we lower the risk of over-fitting. On the contrary, in that case, quadratic fits lead to a slightly better correction of the random walk that limits the stability only up to $\tau \sim 3\times10^4$~s whereas with linear fits, it limits the stability up to $\sim 10^5$~s. Note that, especially for the shortest correction time window we see a clear degradation of the stability for averaging times lower than the correction window's length. This is a known effect from linear servo loop theories and periodic perturbations of oscillators \cite{servo} and it could be attenuated by scaling down the correction: instead of subtracting the result of the fit, we could subtract only a fraction of it. 

Regarding the stability of the corrected Rubidium clock, using linear fits, the conclusions are the same as for the offline correction. The lowest Allan Standard Deviation, for all averaging times, is achieved with the medium width correction time window. With the shortest time window, the short term stability is degraded, and  with the longest correction time window, we find poorer long term stability compared to the other corrected scenarios.

\begin{figure}
    \centering

        \includegraphics[width=\textwidth]{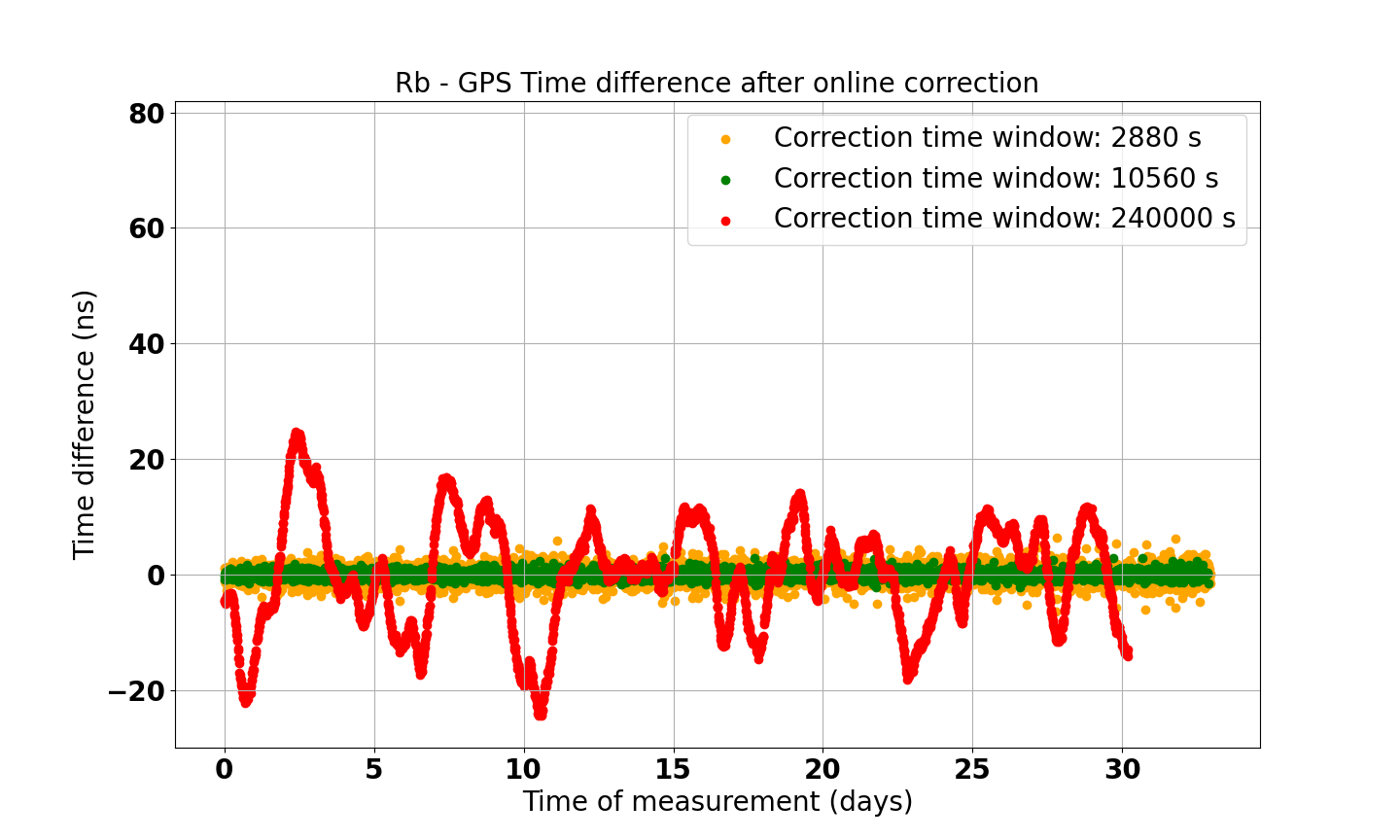}
        \includegraphics[width=\textwidth]{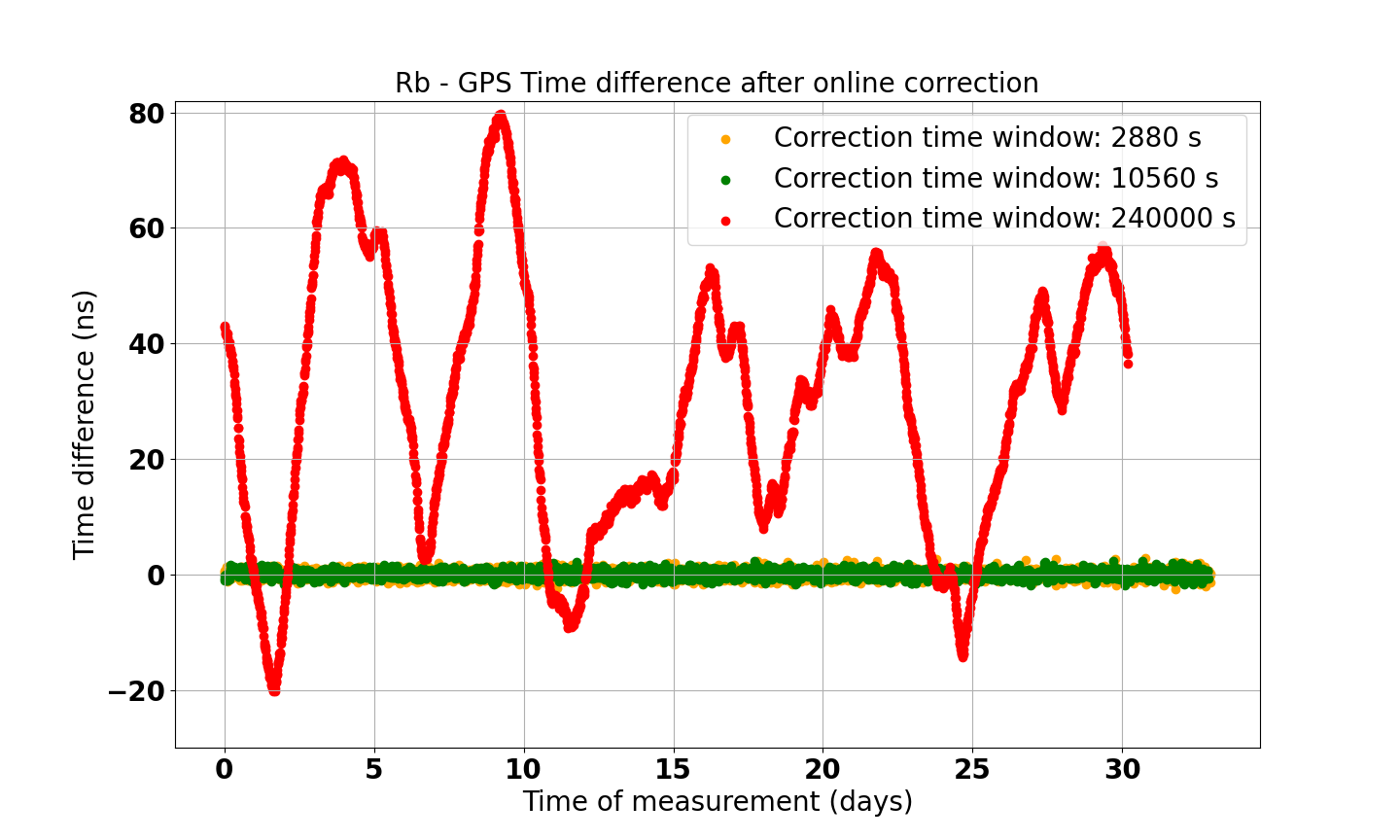}

    \caption{Time difference between the Rubidium clock and GPS Time after the online correction. Each point is corrected using a quadratic (top) or linear (bottom) fit of the $2800$~s (orange) or $10560$~s (green) or $240,000$~s (red) of data points prior to this point. Using linear fits leads to smaller residuals for the shortest time window and bigger ones for the longest time window.}
    \label{fig:residuals_gps_online}
\end{figure}

If the correction time window is too wide, we cannot correct as well the frequency random walk of the free-running Rubidium: the risk is that the Rubidium time signal locally drifts too far away from GPS Time. This can be observed in the corrected Rubidium against GPS Time in Figure~\ref{fig:residuals_gps_online} where the maximum difference reaches around $80$~ns (or $25$~ns with quadratic fits) with the $240,000$~s correction time window. With the $10560$~s correction time window, the differences stay in the $\pm5$~ns range. The standard deviation of the time difference with UTC(OP) is also shown in Table \ref{tab:res_wr} for both online corrections. Once again, one can see the reduction of the white noise when using linear instead of quadratic fits. 
Before correction, as the reader saw in Figure \ref{fig:residuals_beforeCorr}, the free-running Rubidium clock can drift away from GPS Time by around $100$~ns in less than $3$ days which means that HK's requirement for the synchronization with UTC is not met. After online correction with the longest time window tested, the corrected Rubidium time stamps drift by around $60$~ns in a few days because of remaining random walk noise. Even though during the $35$ days data-taking period the time residuals with respect to GPS Time does not exceed $100$~ns, it is not possible to safely claim that the Rubidium clock drift will not exceed HK's requirement of $100$~ns if we use the $240,000$~s correction time window, because of the random nature of this drift. With shorter time windows, no residual drift is observed, and the residuals are thus contained in a range of a few nanoseconds.

\begin{table}[]
    \centering
    \begin{tabular}{c|lll}
        correction time window & $2880$~s &  $10560$~s & $240000$~s\\
        \hline
        offline correction & $1.87$~ns & $1.79$~ns & $5.13$~ns\\
        online correction (quadratic fits) & $2.01$~ns & $1.83$~ns & $9.35$~ns \\
        online correction (linear fits) & $1.84$~ns & $1.81$~ns & $22.66$~ns\\
    \end{tabular}
    \caption{Standard deviation of the time difference between the Rubidium clock PPS signal and UTC(OP) after correction.}
    \label{tab:res_wr}
\end{table}

\section{Discussion}

As advertised before, the advantage of the so-called online correction is that it could be performed in real-time. This is an important feature for applications that necessitate a real-time synchronization with UTC or with another site (like the future HK or DUNE experiments). If a reference clock signal is generated with an atomic clock (like the Rubidium clock used here) and sent to a data acquisition system to be propagated to detectors and provide time stamps, one could continuously compare this signal to GPS Time using a Septentrio receiver. The correction coefficients $a$, $b$ and $c$ calculated from the Septentrio data would need to be sent to the data acquisition system so that it could correct the time stamps in real-time. 

\begin{figure}[tbp]
    \centering
\includegraphics[width=\textwidth]{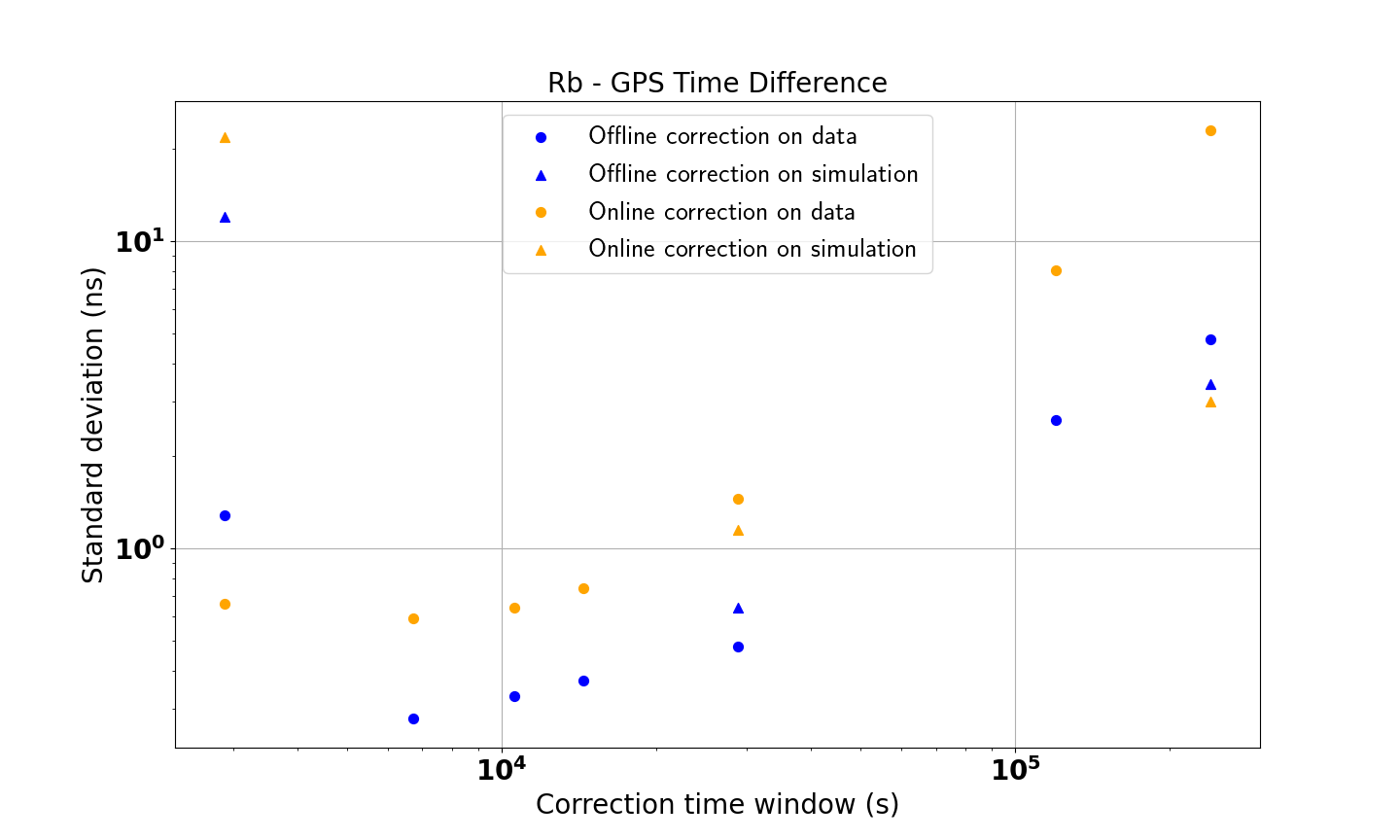}
    \caption{Standard deviation of the residuals distributions between the Rb and GPS Time after the offline (blue) or online (orange) correction as a function of the correction time window. Quadratic fits of the Septentrio data are used for the offline correction whereas linear fits are used for the online correction. The performance on simulated data is also shown for three values of the correction time windows.}
    \label{fig:TimeWindowOpt}
\end{figure}

Figure \ref{fig:TimeWindowOpt} shows the standard deviation of the Rb vs GPS Time difference after correction as a function of the correction time window's width.
The performance of the offline and online corrections on experimental data (colored dots) are compared to the performance we had obtained on simulated data (colored triangles) with a correction time window of $2880$~s, $28800$~s and $240000$~s. Note that these simulated data were only taking into account phase white noise, frequency white noise and frequency random walk components. In particular, the measured data also contain a linear frequency drift and this main difference could partly explain the difference of performances observed between data and simulation. Also, no additional uncertainties were added to take into account other types of noise (e.g: flicker noise) or experimental conditions (e.g: imperfect calibrations, etc.). 
For both corrections, very similar performance of synchronization with GPS Time are obtained for correction time windows below $30,000$~s so there is no need to have much shorter windows. This result is consistent with the fact that, as seen in Figure \ref{fig:ASD_char_Rb}, the stability of the Rubidium signal becomes worse than that of the GPS around $10^4$~s. 
The offline correction seems to provide a slightly better synchronization to GPS Time (down to $\sim 0.3$~ns) but the precision achievable with the online correction is already more than satisfying: better than $5$~ns for correction time windows below $100,000$~s.

\section{Conclusions}

In this paper, we presented a simple way to use time comparisons to GPS Time to synchronize the time stamps, generated using a free-running Rubidium clock, close to UTC while preserving its short term stability and correcting for the long term frequency random walk and deterministic drift. This method has the advantage of using relatively cheap instruments and to be applicable online for a real-time synchronization as well as to be robust against punctual GPS signal reception failures. The online method could be applied for the real-time synchronization between several experimental sites in long-baseline accelerator neutrino experiments as well as for other detectors involved in multi-messenger astrophysics measurements. 

The proposed method consists in fitting GPS Time vs Rubidium measured by a GNSS receiver with a piece-wise polynomial function of time and in subtracting the result to the generated time stamps. The method was first designed and validated with simulated signals before assessing its performance on real data. We evaluated the performance of this correction by quantifying the stability of the clock signal before and after the correction using the Overlapping Allan Standard Deviation.  
We showed that the optimal length of the time window for the fit of GPS Time vs Rubidium seats around $10,000$ seconds, corresponding to around $10$ data points from the receiver. This time window allowed to maintain the best possible short term stability while correcting efficiently the frequency random walk. After correction with this time window, the difference to GPS Time stays within a window of $\pm 5$~ns for both offline and online corrections during the whole period of $35$ days of measurement. This performance largely meets the usual requirements for long-baseline accelerator neutrino experiments, like Hyper-Kamiokande and DUNE. Note that we do not expect the performance of the correction to be heavily degraded by isolated missing or outlier measurements from the receiver. However, this correction requires a constant monitoring of the Rubidium time signal with a GNSS receiver (or other reference that can be linked to UTC). One should thus make sure that such a reference is available in the long term and that there is no risk of loosing it for long periods (e.g.: several hours).

\vspace{6pt}

\paragraph{\textbf{Fundings:}}
This research was funded by IN2P3/CNRS, the French "Agence nationale pour la recherche" under grant number ANR-21-CE31-0008, the "IdEx Sorbonne Université" and the 2019 "Sorbonne Université Émergences: MULTIPLY” grant.\\
The White Rabbit network and the access of associated optical fibers to the Pierre and Marie Curie campus: T-REFIMEVE, FIRST TF and LNE: "Agence Nationale de la Recherche" (ANR-21-ESRE-0029 / ESR/Equipex T-REFIMEVE, ANR-10-LABX48-01 / Labex First-TF);  Laboratoire National d’Essai (LNE), project TORTUE.

% Bibliography

\end{document}